\documentclass[twocolumn]{autart}

\usepackage{graphicx,subfigure}% for pdf, bitmapped graphics files
\usepackage{epsfig} % for postscript graphics files
\usepackage{mathptmx} % assumes new font selection scheme installed
\usepackage{times} % assumes new font selection scheme installed
\usepackage{amsmath} % assumes amsmath package installed
\usepackage{amsmath} 
\usepackage{dsfont}
\usepackage{verbatim}

\usepackage{euscript}

\usepackage{pgfplots}\pgfplotsset{compat=1.4}

\usepackage{xspace}      % use \xspace after macros
\usepackage{amsmath}
\usepackage{bm}
\usepackage{amssymb}
\usepackage{enumitem}
\usepackage{overpic}
\usepackage{hyperref}
\usepackage{graphicx}
\usepackage{cite}        % sorts numbered citations
\usepackage{braket}
\usepackage{subdepth}
\usepackage{color}

\usepackage{pgfplots}\pgfplotsset{compat=1.4}

\definecolor{black1}{rgb}{0,0,0}

\definecolor{softgray}{rgb}{0.92,0.92,0.95}
\definecolor{softblack1}{rgb}{0.90,0.92,1.00}

\definecolor{lightgray}{rgb}{0.12,0.12,0.55}

\definecolor{theframe}{gray}{0.75}
\definecolor{theblue} {rgb}{0.02,0.04,0.48}
\definecolor{thegrey} {gray}{0.5}
\definecolor{theshade}{gray}{0.98}
\definecolor{thered}  {rgb}{0.00,0.00,0.00}
\definecolor{thegreen}{rgb}{0.3,0.3,0.3}

\usepackage{color}
\definecolor{softblue}{rgb}{0.90,0.92,1.00}
\usepackage{mdframed}

\usepackage{graphicx}
\usepackage{hyperref}

\numberwithin{equation}{section}

\newtheorem{lemmer}{Lemma}[section]
\newtheorem{remark}[lemmer]{Remark}

\newtheorem{proposition}[lemmer]{Proposition}

\DeclareMathOperator*{\argmin}{argmin}

\newcommand{\E}{{\mathbb E}}
\newcommand{\Z}{{\mathbb Z}}

\newcommand{\R}{{\mathbb{R}}}

\newcommand{\Var}{\mbox{Var}}
\newcommand{\Trace}{\mbox{tr}}

\newcommand{\F}{\mathcal{F}}

\begin{document}
%TAC GDN su TAC e mostrare domanda abilitazione
\begin{frontmatter}

\thanks[footnoteinfo]{This paper was not presented at any IFAC
meeting. Corresponding author Gianluigi Pillonetto Ph.
+390498277607. This research has been partially supported by the MIUR FIRB project RBFR12M3AC-Learning
meets time: a new computational approach to learning in dynamic
systems, by the Progetto di Ateneo CPDA147754/14-New statistical learning approach for multi-agents adaptive estimation and coverage control  
as well as by the Linnaeus Center CADICS, funded by the
Swedish Research Council, and the ERC advanced grant LEARN, no 287381, funded by the European Research Council.}

\title{Regularized linear system identification using 
atomic, nuclear and kernel-based norms: the role of the stability constraint}

\author[First]{Gianluigi Pillonetto}
\author[Second]{Tianshi Chen}
\author[First]{Alessandro Chiuso}
\author[Third]{Giuseppe De Nicolao}
\author[Second]{Lennart Ljung}

\address[First]{Department of Information  Engineering, University of Padova, Padova, Italy (e-mail: \{giapi,chiuso\}@dei.unipd.it)}
\address[Second]{Division of Automatic Control, Link\"oping University, Link\"oping, Sweden (e-mail: \{tschen,ljung\}@isy.liu.se)}
\address[Third]{Department of Computer Engineering and Systems Science, University of Pavia, Pavia, Italy (e-mail: giuseppe.denicolao@unipv.it)}


\maketitle
\begin{abstract}
Inspired by ideas taken from the machine learning literature, new regularization techniques have been recently introduced in linear system identification. In particular, all the adopted estimators solve a regularized least squares problem, differing in the nature of the penalty term assigned to the impulse response. Popular choices include atomic and nuclear norms (applied to Hankel matrices) as well as norms induced by the so called stable spline kernels. In this paper, a comparative study of estimators based on these different types of regularizers is reported. Our findings reveal that stable spline kernels outperform approaches based on atomic and nuclear norms since they suitably embed information on impulse response stability and smoothness. This point is illustrated using the Bayesian interpretation of regularization. We also design a new class of regularizers defined by ``integral" versions of stable spline/TC kernels. Under quite realistic experimental conditions, the new estimators outperform classical prediction error methods also when the latter are equipped with an oracle for model order selection. 
\end{abstract}

\end{frontmatter}

\section{Introduction}

Prediction error methods (PEM) are a classical tool for reconstructing the impulse response
of a linear system starting from input-output measurements  \cite{Ljung:99}. 
In the simplest scenario, one postulates a single model structure, e.g. a rational transfer
function which depends on an unknown parameter vector $g$ of known dimension.  
If the model contains the ``true" system, and the noise source is Gaussian, estimation of $g$
by PEM enjoys optimal asymptotic properties. In particular, this estimator can not be 
outperformed by any other unbiased estimator as the number of measurements goes to infinity.\\
However, in real applications, not only the number of measurements is always finite but also model complexity is typically unknown. This means that different model structures need to be introduced, e.g. rational transfer functions of different order. Each of them has to be fitted to data by PEM and then compared resorting e.g. to validation techniques (cross validation) or complexity criteria such as Akaike's criterion (AIC) \cite{Efron04,Akaike:74}.
Recent studies have however illustrated some limitations of this approach. For instance,
when the data set size and/or the signal to noise ratio is not so large, it can return models 
with a non satisfactory prediction power on new data \cite{SurveyKBsysid}.\\

The above issues have motivated the development of an alternative route to system identification
based on regularization techniques. The starting point is the use of high-order FIR models in combination with penalty terms (regularizers) on the impulse response $g$. 
In particular, an important class of estimators solves a convex optimization problem of the form 
$$
\arg\min_g \ V(g)+\gamma J(g), \quad \gamma \geq 0.% \in \mathcal{R}^+ 
$$
Above, $V$ is the so-called loss function that 
depends also on the input-output measurements and measures the adherence to experimental data. 
It is often given by the sum of squared residuals (the choice also adopted in this paper), leading to regularized least squares (ReLS).
The term $J$ is instead the regularizer which typically includes smoothness information on $g$. 
Finally, the positive scalar $\gamma$ is the
regularization parameter which has to suitably balance $V$ and $J$. In practice, it is always unknown and has to be determined from data.\\

An important advantage of ReLS over classical PEM is that the difficult model order determination can be replaced by estimation of few regularization variables. For instance, the stable spline estimator introduced in \cite{SS2010} depends only on two parameters: the regularization parameter $\gamma$ and another one which enters $J$ by determining
how fast $g$ is expected to decay to zero. Such variables can be determined e.g. by cross validation \cite{Efron04}, $C_p$ statistics \cite[subsection 7.4]{Hastie01} or marginal likelihood optimization \cite{Maritz:1989,MacKay}, an approach recently proved to be much effective \cite{SS2010,ChenOL12,MLECC2014}.
In particular, in the spirit of Stein's effect \cite{JamesStein1961}, a carefully tuned regularization can much reduce the variance of the estimates just introducing a small bias in the identification process. Hence, the mean squared error of the estimator can turn out inferior than that achieved by PEM \cite{SS2010,ChenOL12}. However, to obtain this, the choice of $J$ is crucial since it has a major effect on the quality of impulse response reconstruction. It is thus of interest to investigate and compare the performance of different regularizers recently proposed in the system identification literature. 

A recent regularization approach relies on the so called {\it{nuclear norms}}.
The nuclear norm (or trace norm) of a matrix is the sum of its singular values.
Being  the convex envelope of the rank function on the spectral ball, it has been often used as a convex surrogate of the rank function \cite{Fazel01}. 
In particular, since the McMillan degree (minimum realization order) of a discrete time time-invariant system coincides with the rank of the Hankel operator constructed from the impulse response coefficients, 
it is tempting to set $J$ to the nuclear norm of the Hankel operator associated with the impulse response $g$.
In this way, output data are described   
while encouraging a low McMillan degree, see \cite{liu2009,GrossmanCDC09,Mohan2010,Smith2014} and also \cite{Signoretto2012,Hjalmarsson2012}
for extensions of this idea for estimation e.g. of Box-Jenkins models.
%An atomic norm regularizer \cite{Chand2012} which mimics the Hankel nuclear norm, and is approximated by
%a suitable $\ell_1$ penalty, is described in \cite{Shah2012}.

{\it{Atomic norms}} have been also adopted as regularizers for system identification in the last years \cite{Chand2012}. The function to be reconstructed is described as the sum of a (possibly infinite) number of basis functions, dubbed {\it{atoms}}. The penalty $J$ is then given by the atomic norm defined, under some technical conditions, by the convex hull of the atomic set.  For instance, the convex hull of one-sparse vectors of unit Euclidean norm leads to the $\ell_1$ norm which enjoys important recovery properties \cite{Donoho2006a}, being also related to the popular LASSO procedure \cite{Tibshirani94}.\\
A motivation underlying the use of the atomic norm is that, in some sense, it represents the best convex penalty when the function is sum of few atoms. Successful applications in signal processing and machine vision regard estimation of sparse vectors and low-rank matrices \cite{Donoho2006b,Cands2009,Aja2009}. This technique has been recently introduced also in the system identification scenario in \cite{Shah2012} exploiting low-order rational transfer functions as atomic set, see also \cite{Rojas1,Rojas2014} for other approaches that use the $\ell_1$ penalty. %[]altri approcci l1

Other recent system identification techniques exploit penalty terms induced by {\it{kernel}} functions
\cite{Scholkopf01b,Bottou07}, which have to capture the expected features of the unknown function.
In particular, the works \cite{SS2010,SS2011,ChenOL12}  have proposed a class of kernels for system identification, 
including {\it{stable spline (SS)/tuned-correlated (TC)}}, which encodes information on smoothness and exponential stability of $g$, 
see also \cite{ChiusoCLPCDC2014} for other insights on the stable spline kernel structure.

The goal of this paper is to compare the performance of ReLS
equipped with atomic, nuclear or kernel-based norms via numerical studies.
For this purpose, we will perform a 
Monte Carlo experiment where at every run the true system is a different 
rational transfer function randomly chosen by a MATLAB generator.
The system input and the signal to noise ratio also 
varies from run to run, thus leading to a large variety of system identification problems.
The results reveal that, in many cases, atomic and nuclear norms 
lead to unsatisfactory impulse response estimates. 
The drawbacks of these approaches are explained through the Bayesian interpretation 
of regularization, where different $J$ are seen as  
different a priori probability density functions (pdf) assigned to $g$.  
This interpretation explains why the current implementations of 
atomic and nuclear norms fail in capturing 
important characteristics of stable dynamic systems.  
As far as the nuclear norm is concerned, when used in conjunction with a FIR model of order $m$,
we shall see that it essentially corresponds to modeling 
the impulse response coefficients $g_t$ as a non stationary white noise process with a 
variance roughly decaying as $1/t$ for $t<m/2$ and increasing as $1/(m-t)$ for $t>m/2$.
%For what regards the Hankel nuclear norm, quite surprisingly, we will see that it essentially
%corresponds to modeling the impulse response coefficients as white noise 
%with variance first decreasing and then increasing. 
This means that the prior underlying this approach does not embed any information on impulse response stability 
and smoothness, two key features to achieve good mean squared error properties 
\cite{SurveyKBsysid}. As for ReLS equipped with the atomic norm 
described in \cite{Shah2012}, we will see that it suffers of the
limitations of the LASSO procedure recently discussed in \cite{AravkinJMLRconv}.
%\\connecting  such estimator with 
%the LASSO
%with an approach called {\it{multiple kernel learning}} in the machine learning literature \cite{MKL2011}, 
%some drawbacks of the LASSO procedure recently discussed in \cite{AravkinJMLRconv}.\\

The results from the same Monte Carlo study will also show that estimators
based on stable kernels 
outperform ReLS relying on atomic and Hankel nuclear norms. 
Furthermore, we will design a new class of regularizers induced by
``integral" versions of stable spline kernels. 
This will lead to novel ReLS approaches for system identification 
that may outperform also the classical PEM equipped with an oracle for model complexity selection 
(the oracle selects the best model order exploiting information on the true system).
%The reasons 
%will still be  explained in a Bayesian framework, proving that the
%stable spline kernel introduced in in \cite{SS2010,PillACC2010} enjoys Maximum Entropy (MaxEnt) properties.
%This means that, in some sense, it represents   
%the least committing prior
%when smoothness and stability is the sole information on $g$. 
%These results will be also connected with atomic norm approaches,
%by deriving the concepts of MaxEnt atomic set and MaxEnt regularizer.
%The same analysis will also motivate the 
%design of a new class of regularizers induced by
%``integral" versions of stable spline kernels. 
%This will lead to novel ReLS approaches for system identification 
%that may outperform also the classical PEM equipped with an oracle for model complexity selection 
%(the oracle selects the best model order exploiting information on the true system).

The paper is so organized. Section \ref{Sec2} reports the problem statement
while in Section \ref{Sec3} we briefly review the ReLS
estimators based on the Hankel nuclear norm and the atomic norm. 
Section \ref{Sec4} gives insights about
limitations of the techniques described in Section \ref{Sec3} using the Bayesian
interpretation of regularization. In Section \ref{Sec5} we briefly review
stable spline kernels introduced in \cite{SS2010,PillACC2010}, also introducing the concepts
of stable spline atomic set. In Section \ref{Sec6}, new integral versions of stable spline kernels
are introduced and used to define novel ReLS estimators
for linear system identification. All of the estimators are then tested via Monte Carlo studies
in Section \ref{Sec7}. Conclusions end the paper while some mathematical details
are gathered in Appendix.

\section{Problem statement}
\label{Sec2}

We use $u(t)$ and $y(t)$ to denote, respectively, the input and the noisy output 
of a SISO system observed at time instant $t$. For convenience of notation we shall also use the notation $y_i:=y(t_i)$. 
The measurement model we consider is \color{black1} of the Output-Error (OE) type: \color{black}
\begin{equation}
\label{MeasMod}
y_i = (g \otimes u)_i +e_i, \quad i=1,\ldots,N %   (g \otimes u)_i +e_i, \quad i=1,\ldots,n
\end{equation}
where $g$ is the system impulse response,
$(g \otimes u)_i$ denotes the convolution in discrete time between $g$ and $u$ evaluated at $t_i$.
Finally, $e_i$ are independent Gaussian noises of
constant variance $\sigma^2$. \color{black1} The results  could be straightforwardly extended
to non stationary noises and/or ARMAX/BJ type structures.   Our problem is to estimate the impulse response $g$
assuming the input $u$ is (deterministic and) known and having collected the measurements $Y=[y_1 \ \ldots \ y_N]^T$.
 \color{black} 
\color{black1}
\begin{remark}
In the sequel, we will discuss regularized estimators based on IIR (infinite impulse response) or FIR (finite impulse response) models for $g$ and equipped with a regularizer $J(g)$. 
In the IIR case, following the well known concept of BIBO stability, 
$J$ is said to include the stability constraint if it embeds the constraint $\sum_{t=1}^\infty |g_t| < \infty$.
FIR models are already structurally stable. However, also in this case 
we will often use expressions as {\emph{lack of stability constraint}} 
to mean that $J$ does not include the information 
that impulse response is expected to decay to zero as a function of the time lag. 
In particular, this concept will be formalized in Bayesian terms.
\end{remark}
\color{black}

\section{ReLS based on Hankel nuclear/atomic norms}
\label{Sec3}

In this section we describe two ReLS approaches which use as regularizer $J$ \color{black1}
either the Hankel nuclear norm or a version of the atomic norm that can be approximated by the 
$\ell_1$  norm. \color{black}%Both of the techniques are implemented in discrete-time. 

\subsection{Regularization via Hankel nuclear norm}\label{Hankel}

We introduce a regularizer based on the Hankel nuclear norm.
Recall that the nuclear norm of a matrix $A$ is the sum of its singular values, i.e.
$$
\|A\|_* = \sum_i \sigma_i(A).
$$
The motivation underlying the use of this type of penalty for system identification stems from the 
fact that the minimum realization order (also known as McMillan degree) of a discrete time LTI system coincides with the rank of the Hankel operator $H(g)$ constructed from the impulse response coefficients $g_k$, i.e.
\begin{equation}\label{HankM}
H(g) = \left(
         \begin{array}{cccc}
           g_1 & g_2 & g_3 & \cdots \\
           g_2 & g_3 & g_4 & \cdots  \\
           g_3 & g_4 & g_5  & \cdots \\
           \vdots &  \vdots   &  \vdots  & \ddots\\
         \end{array}
       \right).
\end{equation}
Then, as described e.g. in \cite{GrossmanCDC09,Mohan2010}, an 
estimator trading data fit with low McMillan degree can be obtained by solving
\begin{equation}\label{HankelEst}
\arg \min_{g}  \ \sum_{i=1}^N \left( y_i - (g \otimes u)_i \right)^2 + \gamma \| H(g) \|_*
%\arg \min_{g} \| Y - \Phi g \|^2 + \gamma \| H(g) \|_*
\end{equation}

\subsection{Regularization via atomic norm}\label{subsecRAN}

A different, but related regularization approach to linear system
identification, called the atomic norm regularization approach, was
recently suggested in \cite{Shah2012}. It describes the model using ``atoms'' and defines a model complexity measure in terms of the
``atomic norm''. This complexity measure is then used for regularization.

Let $\mathcal{C}$ be the complex plane and $\mathcal{D} = \left\{ w
\in \mathcal{C}, \ |w| < 1  \right\}$ and consider below the
discrete-time case. A simplistic account of the idea is as follows:
Given a set of ``atoms'',  $G_w(z), w\in\mathcal{D}$ (which can be
interpreted as basis functions for the model transfer function), we
can construct a linear model via linear combinations of the atoms. For
a concrete feeling of the concept, think of the atoms as normalized
first order system with pole (possibly
complex) denoted by $w\in \mathcal{D}$, so
\begin{equation}\label{Atset}
G_w(z)=\frac{1-|w|^2}{z-w}.
\end{equation}
\color{black1} The transfer function of any linear system  can be written \color{black} as a finite or
countable linear combination of these atoms:
\begin{equation}\label{AtomicDec}
G(z) = \sum_k a_kG_{w_k}(z),\quad w_k\in\mathcal{D}.
\end{equation}
The atomic norm of this system is defined as
\begin{align} \label{Atomic}
\|G(z)\|_{\mathcal{A}} = \inf\left\{\sum_{w_k\in\mathcal{D}}
|a_k|  \  :  \  (\ref{AtomicDec}) \ \mbox{holds}  \right\}.
\end{align}
It has been proved in \cite{Shah2012} that the atomic norm well
approximates the Hankel nuclear norm of $G(z)$, thus motivating its
use in system identification.

For computational reasons, a finite number of atoms $G_{w_k}(z)$,
$k=1,\cdots,p$ are selected. If $G_{w_k}(z)$ is selected and
$w_k\in\mathcal{C}$, then $G_{w_k^*}(z)$ shall be also within the
selected $p$ atoms where $w_k^*$ is the complex conjugate of $w_k$.
Let the impulse response of $G_{w_k}(z)$ be $\rho_{w_k}$ and denote
$$h_{w_k}(i)= (\rho_{w_k} \otimes u)_i$$
%$$h_{w_k}(t)= \sum_{i=1}^{t-1} \rho_{w_k}(i)u(t-i).$$
\color{black1}
Then, recalling (\ref{AtomicDec}), 
\color{black} the fit between the measured output and the model is
$$\sum_{i=1}^N \left(y_i-\sum_{k=1}^p a_kh_{w_k}(i)\right)^2$$
%$$\sum_{t=1}^N |y(t)-\sum_{k=1}^p a_kh_{w_k}(t)|^2$$ %\sum_{t=1}^N |y(t) -G(q)u(t)|^2 = 
and the atomic norm regularized estimate becomes the LASSO like
\begin{equation} \label{Atomic2}
\hat a = \argmin_a \sum_{i=1}^N \left(y_i-\sum_{k=1}^p a_kh_{w_k}(i)\right)^2 + \gamma
\sum_{k=1}^p |a_k|.
\end{equation}
%\begin{equation} \label{Atomic2}
%\hat a = \argmin_a \sum_{t=1}^N |y(t)-\sum_k a_kh_{w_k}(t)|^2 + \gamma
%\sum_{k=1}^p |a_k|.
%\end{equation}

\section{Bayesian interpretation of regularization: the Hankel nuclear norm and
the atomic norm case}
\label{Sec4}

Every ReLS estimator
\begin{equation}\label{BeforeMAP}
\arg \min_{g} \ \sum_{i=1}^N \left( y_i - (g \otimes u)_i \right)^2  + \gamma J(g)
\end{equation}
can be given a simple interpretation in Bayesian terms.
To simplify the exposition, %consider the discrete-time setting and
assume that the measurements model (\ref{MeasMod}) 
can be written as a FIR of order $m$. Then, abusing notation,
we use $g$ to denote the (column) vector containing the
impulse response coefficients $g_1,\ldots,g_m$. More specifically, assume 
that $g$ is a random vector with pdf
\begin{equation}\label{BayesJ}
p(g) \propto  \exp\left(-\frac{J(g)}{2\lambda}\right)
\end{equation}
\color{black1} where ``$\propto$" stands  for ``proportional to" while $\lambda$ is a 
positive scale factor. \color{black}  Assume also that
$Y$ is corrupted by additive zero mean white Gaussian noise, 
independent of $g$, of variance $\sigma^2$, i.e.
$$
p(Y | g) \propto  \exp\left(-\frac{\sum_{i=1}^N \left( y_i - (g \otimes u)_i \right)^2 }{2 \sigma^2}\right).
$$ 
By the Bayes rule, the posterior of $g$ is the product of the likelihood and the prior, i.e. $p( g | Y) \propto p(Y | g) p(g)$,
so that
$$
-\log p(g | Y) = \frac{\sum_{i=1}^N \left( y_i - (g \otimes u)_i \right)^2 }{2\sigma^2} + \frac{J(g)}{2\lambda} + \mbox{Const.}
$$
Hence, setting $\gamma=\sigma^2/\lambda$, one can conclude that ReLS (\ref{BeforeMAP}) can be
always seen as a maximum a posteriori (MAP) estimator.%[]ref  

The estimator (\ref{BeforeMAP}) obtained  from this Bayesian perspective  will perform well only provided the density $p(g)$, 
defined through (\ref{BayesJ}), succeeds in capturing the essential features and possibly prior knowledge on 
the underlying dynamical system. In particular 
the prior should privilege stable and possibly smooth impulse responses.
Now, it is of interest to  elucidate  
the shape of the priors underlying the two ReLS approaches introduced in the previous section.
%For our analysis, the following well known result will be useful \cite{Andrews}: 
%if $x$ is a Laplacian random variable of mean $0$ and variance $2\lambda^{-2}$, its pdf
%can be written as a mixture of normals with an exponential mixing density, i.e.
%\begin{equation}\label{BasicMix}
%p(x) = \frac{1}{2 \lambda} e^{-\frac{|x|}{\lambda}} = \int_0^{+\infty} \frac{1}{\sqrt{2 \pi} \nu} e^{-\frac{y^2}{2\nu^2}} \frac{1}{2\lambda^2} e^{-\frac{\nu^2}{2 \lambda^2}}dy.
%\end{equation}

\subsection{The Hankel nuclear norm case}\label{HankelPrior}

\begin{figure*}
  \begin{center}
   \begin{tabular}{cccc}
\hspace{.1in}
 { \includegraphics[scale=0.4]{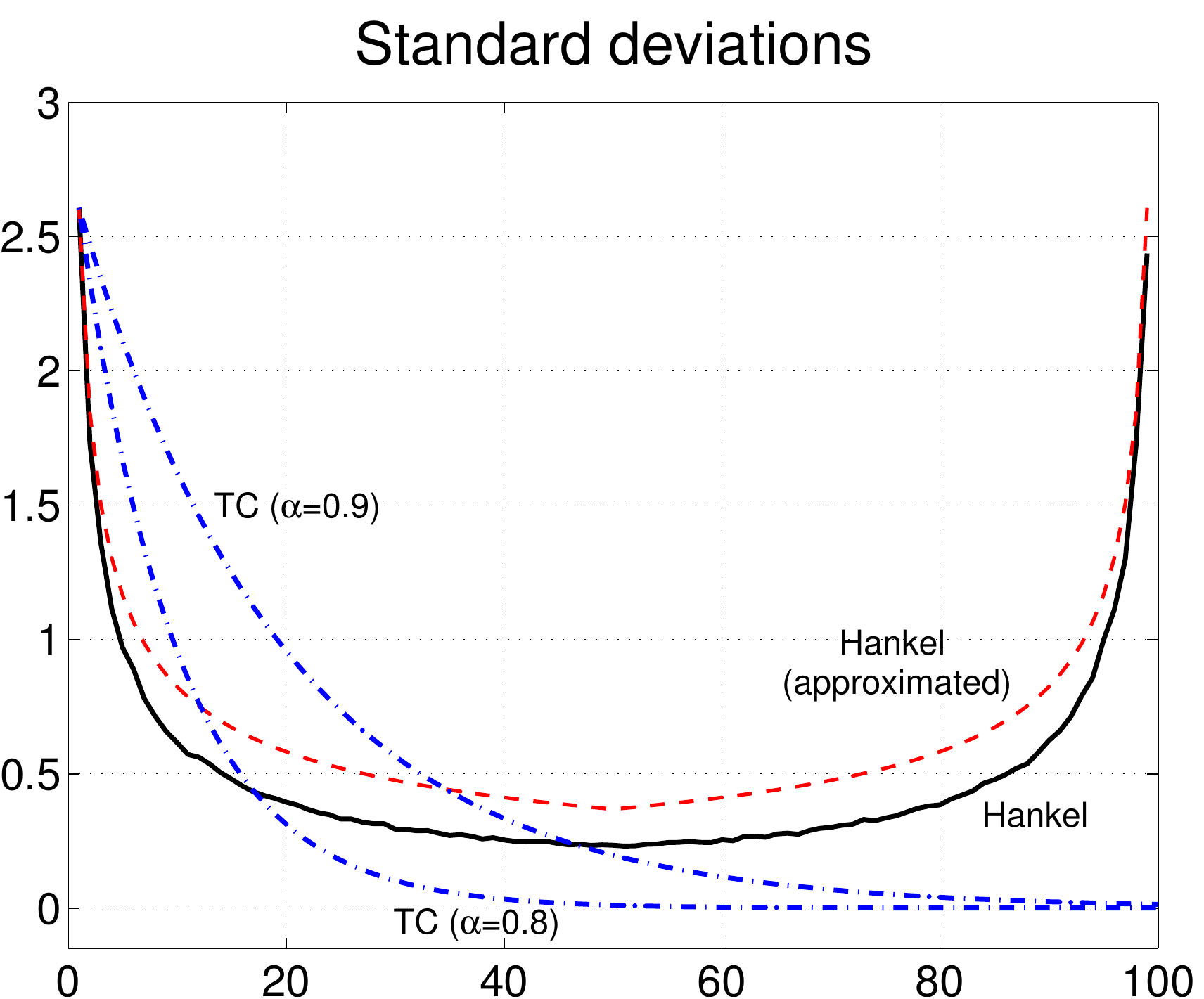}} 
\hspace{.1in}
 { \includegraphics[scale=0.4]{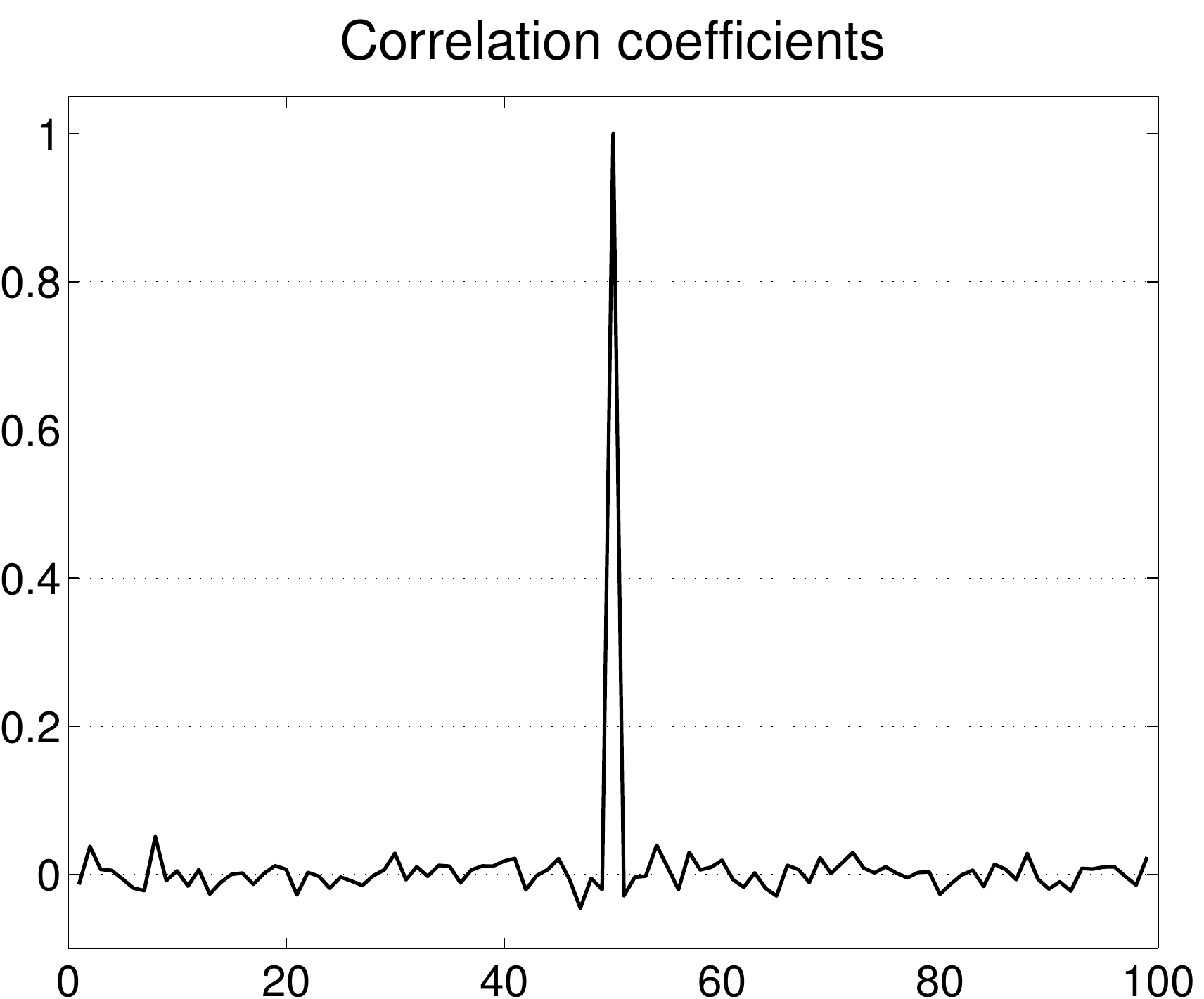}} 
    \end{tabular}
 \caption{{\bf{Prior induced by the Hankel Nuclear Norm}}: the impulse response coefficients are contained in the 
 vector $g \in \R^{99}$, modeled
as a random vector with probability density function $p_H(g) \propto \exp(-\| H(g) \|_*)$. Here, $\| \cdot \|_*$ is the  nuclear norm while $H(g)$ is the Hankel matrix (\ref{HankM}) of size $50 \times 50$. {\it{Left}}: standard deviations of the impulse response coefficients $g_k$ reconstructed by MCMC (solid line) and approximated using the prior (\ref{Htilde},\ref{pHapprox}) (dashed line)
derived in Appendix. The figure also displays the standard deviations of $g_k$ when $g$
is a Gaussian random vector with stable spline covariance (\ref{SS1}) for two different values of 
$\alpha$ (dashdot lines). 
  {\it{Right}}:  50-th row of the matrix containing the correlation coefficients returned by the MATLAB command \texttt{corrcoef(M)} 
  where each row of the 1e6$\times 99$ matrix \texttt{M} contains one MCMC realization of $g$ under the Hankel prior $p_H(g) \propto \exp(-\| H(g) \|_*)$.} 
    \label{Fig3}
     \end{center}
\end{figure*}

%Let us come back to the nuclear norm regularizer. 
According to the previous Bayesian interpretation of regularization, 
the nuclear norm regularizer
\begin{equation}\label{HankelNuclearNorm}
J(g) = \sum_i \sigma_i(H)
\end{equation}
is associated with the prior
$$
p_H(g)  \propto  \exp\left({-\frac{\sum_i \sigma_i(H)}{2\lambda}}\right).
%p_H(g) \propto \exp\left(-\frac{\|H\|_*}{\lambda}\right), \quad \|H\|_* = \sum_i \sigma_i(H) % = C e^{- \sum_i \sigma_i(H)} }
$$
Now, we would like to understand how well $p_H$
describes typical features of an impulse response.
\color{black1}
By symmetry, it is easy to assess that $g$ is zero-mean but
then a simple closed form expression for the distribution of its components $g_k$ 
is not available. \color{black} 
\color{black1}  In Appendix \ref{App:Hankprior} we describe a Markov chain Monte Carlo (MCMC) scheme  to efficiently sample from  $p_H$  \cite{Gilks}, also exploiting a close form approximation of   
 of $p_H$. \color{black} 
Fig. \ref{Fig3}  reports some results extracted from
a chain of length 1e6 setting $2\lambda=1,g \in \R^{99},H \in \R^{50 \times 50}$. The left panel  
shows the standard deviations of the impulse response coefficients $g_k$ 
(solid line) while the right panel displays information on the correlation
between the components of $g$. 
The outcomes show that, under $p_H$, 
the impulse response coefficients $g_k$ are (approximately) uncorrelated and with bathtub 
variance. This suggests that the prior associated with the Hankel nuclear norm
provides a weakly informative guess when searching for stable and smooth impulse responses.
This analysis also shows that even an apparently reasonable regularizer could be associated to
a meaningless prior distribution.

\begin{remark}\label{rem:hanker:stability}
Lack of stability, manifesting itself in the fact that the impulse response \color{black1} prior \color{black} variance does not decay to zero as a function of the \color{black1} lag \color{black} index $t$, should not come as a surprise. In fact, for any fixed $m$  and any given ``stable'' partial impulse response $\{g_t\}_{t=1}^m$, it is always possible to find an unstable impulse response $\{h_t\}_{t=1}^m$ with the same Hankel nuclear norm. A simple example goes as follows:  consider the (finite) Hankel matrix $H_m(g)$ built with the first $m$ impulse response coefficients of  $g_t = ca^{t-1}b$ and the matrix $H_m(h)$ built with $h_t = a^{m-1}c\left(\frac{1}{a}\right)^{t-1}b$. It is simple to check that $H_m(g)$ can be obtained from $H_m(h)$ by reversing the order of rows and columns and thus $\|H_m(h)\|_*=\|H_m(g)\|_*$. 
\end{remark}

\color{black1}
\begin{remark}\label{rem:hanker:MSE}
In \cite{ChenOL12}, working under a deterministic framework, it has been shown that
favorable mean squared error (MSE) properties in the reconstruction of stable exponentials can be obtained letting 
the diagonal (and off-diagonal) elements of the regularization matrix  
decay exponentially to zero (as the entry number increases). 
In the Hankel case, it is shown in Appendix \ref{App:Hankprior} that one approximately has $\Var(g_t) \propto t^{-1}$ for  $1\leq t \leq \frac{m+1}{2}$   and $\Var(g_t) \propto (m-t+1)^{-1}$ for $ \frac{m+1}{2}< t \leq m $. So, the Hankel regularizer does not include exponential dynamics, a key point 
to well trade-off bias and variance. Note also that the 
particular profile of $\Var(g_t)$ induced by Hankel also implies that the choice of $m$ 
has an important effect on the structure of the Hankel norm regularizer.
\end{remark}
\color{black}

\begin{figure*}
  \begin{center}
   \begin{tabular}{cccc}
\hspace{.1in}
 { \includegraphics[scale=0.4]{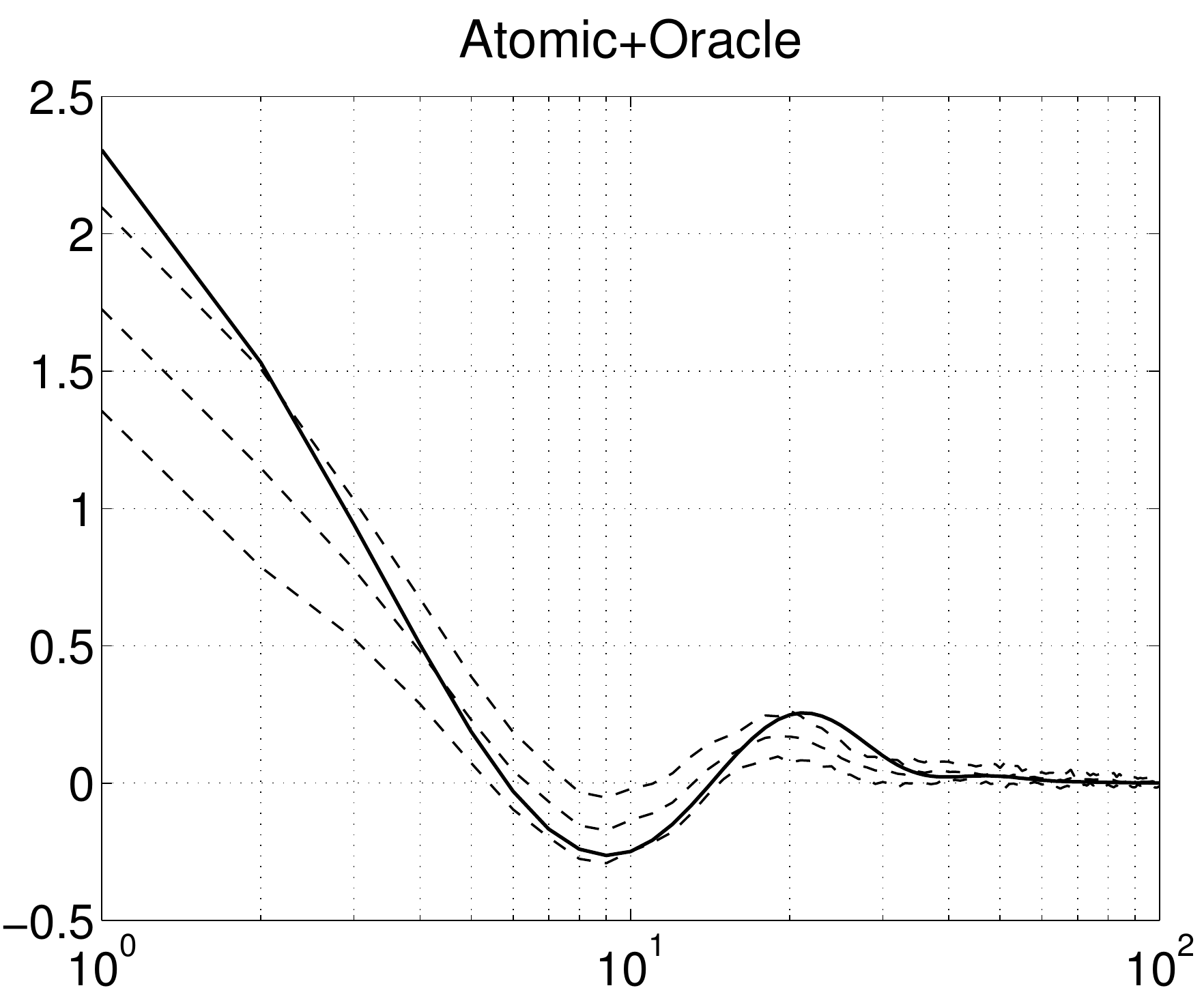}} 
\hspace{.1in}
 { \includegraphics[scale=0.4]{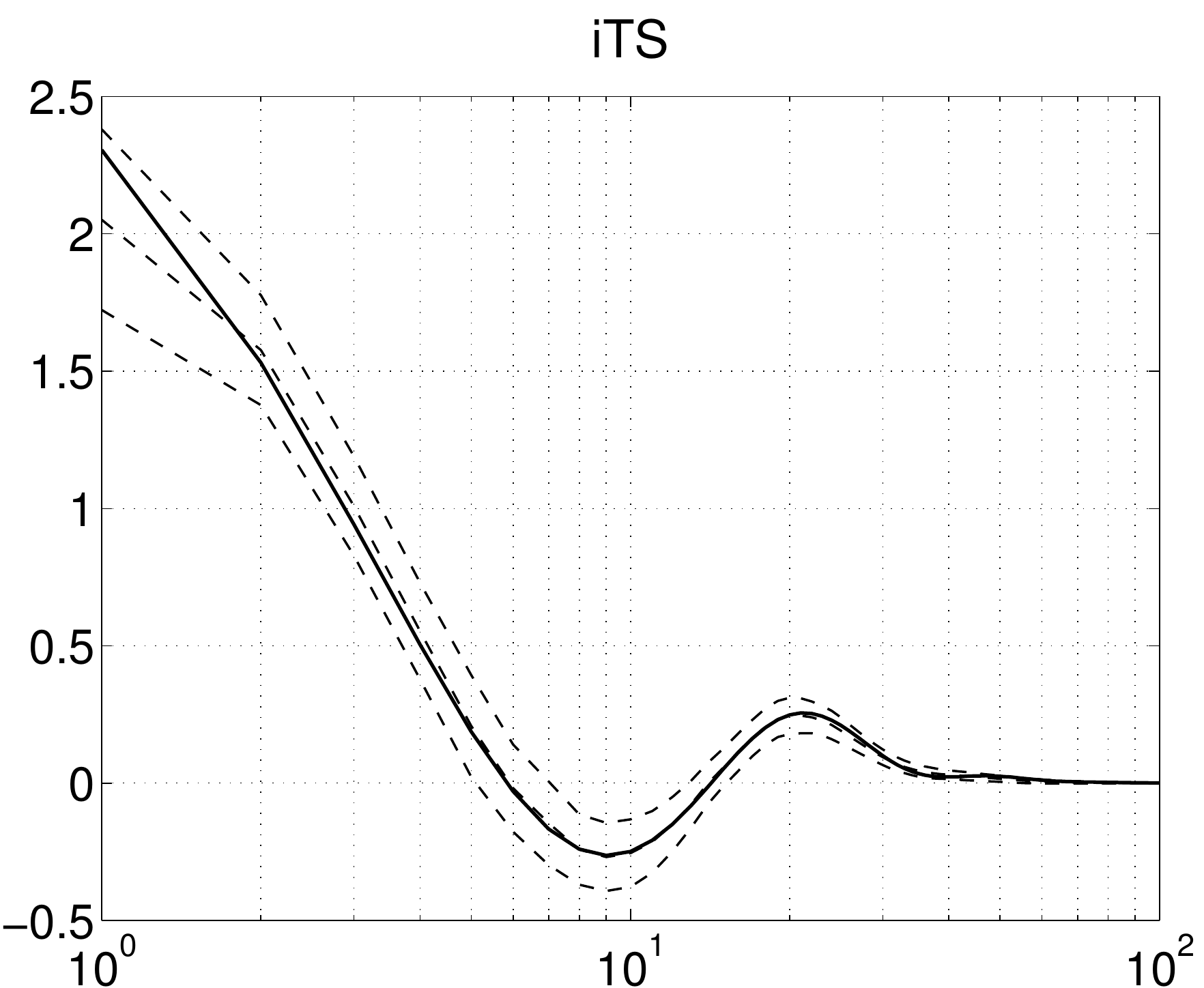}} 
    \end{tabular}
 \caption{{\bf{Monte Carlo experiment with a fixed impulse response:}} true impulse response 
 (thick line) and mean$\pm$standard deviation (dashed line, computed after 100 runs) of the estimators  At+Or (left) and  iTS
(implemented as described in subsection \ref{SevEst}). Recall that, differently from  iTS,  At+Or is not implementable in practice since 
 $\gamma$ is tuned using the true impulse response.} 
    \label{FigFix}
     \end{center}
\end{figure*}

\subsection{The atomic norm case}
\label{secANprior}
%The fact that the atomic norm used in (\ref{Atomic2})
%approximates the Hankel  norm, see \cite{Shah2012},
%is already an indication that the related prior may suffer the same drawbacks 
%described in the previous case. 
\color{black1}  It has been shown in \cite{Shah2012} that, restricting to $\{g_t\}_{t\in \Z^+}  \in \ell_2$, the atomic norm (\ref{Atomic2})
is equivalent to the Hankel   nuclear norm.  For this equivalence to extend to the ``finite''  nuclear norm \eqref{HankelNuclearNorm}  the size of the Hankel matrix \eqref{HankM} (and thus the truncation index $m$) needs to grow to infinity.   Note that the restriction $\{g_t\}_{t\in \Z^+}  \in \ell_2$ is critical since, as we have seen, for any finite $m$ the Hankel nuclear norm does not include  a ``stability'' constraint.  The situation is  different with the atomic norm as any finite sum of (stable) atoms \eqref{AtomicDec} will always result in a stable impulse response.  \color{black}  Yet the $\ell_1$ penalty on the coefficients in \eqref{Atomic2} (see also \eqref{AtomicNorm2} below) may introduce severe bias. More insights are discussed below.\\
Let $g_t = \sum_{j=1}^{p} a_j \rho_j(t)$.
In view of (\ref{Atomic2}), the atomic norm regularizer 
\begin{equation}\label{AtomicNorm2}
J(g) =  \sum_{j=1}^{p} | a_j | %, \ \mbox{for} \ g_t = \sum_{j=1}^{p} a_j \rho_j(t) 
\end{equation}
amounts to a Bayesian prior
\begin{equation*}\label{ANprior}
%\begin{center} p_H(g) \approx 
p_{AN}(g) \propto  \exp\left({-\frac{\sum_{j=1}^{p}  |a_j|}{2\lambda}}\right) 
\end{equation*}
so that the unknown parameters $a_j$ are Laplace distributed independent random variables,
all having the same variance.
As in the case of LASSO, the solutions of (\ref{Atomic2}) enjoy some sparsity properties
meaning that, when $\gamma$ increases, more and more elements of vector $a$ 
are forced to zero.  This may seem an appealing property but the results may fail to \color{black1}  meet the
expectations\color{black}. In fact, the $\ell_1$ penalty may introduce an excessive penalty
on some large coefficients $a_j$ to obtain sparsity, as \color{black1} documented in the Statistics literature \cite{FanLiSCAD_JASA2001,AdaptiveLasso} and also recently demonstrated and discussed in \cite{AravkinJMLRconv}.  \color{black} 
In particular, assume that the true impulse response $g$ is the sum of a finite number of  atoms.
For a large enough value of $\gamma$, all the $a_j$ which do not contribute to $g$
will go to zero but the linear penalty in  (\ref{Atomic2}) will still yield to biased estimates of the other expansion coefficients, so that the resulting estimate is oversmoothed (biased). 
An example of this phenomenon is
illustrated in Fig. \ref{FigFix} (left panel) which reports the results
obtained  via a Monte Carlo study where the impulse response
$g$ \color{black1}  is fixed \color{black}  while different noise and input realizations are generated at every run, 
in the same way as described in Section \ref{Sec7}. The estimates  of $g$ are obtained by (\ref{Atomic2}) using 300 input-output samples, with $\gamma$ chosen by an oracle. The latter has access to the true $g$ 
and selects the regularization parameter which minimizes the mean squared error 
(the concept of oracle-based estimation is further detailed in subsection \ref{SevEst}). 
Under a Bayesian viewpoint, the estimator suffers from the equal probability assigned to all the atomic functions
so that the oracle can select few atoms only assuming that the prior variance (which is proportional to $\lambda$) 
is quite low. Obviously, this introduces a bias. % that explains also the results in Fig. \ref{Fig1}. 
It is worth asking if the adoption of different atoms and
an unequal weighting on their expansion coefficients may be a remedy.
We explore this issue in the next section. %[]cita JMLR

\section{Stable spline kernels and atoms}
\label{Sec5}

In the first part of this section we review the stable spline kernel, comparing its features with the prior
induced by the Hankel nuclear norm discussed in section \ref{HankelPrior}. We then 
compare the structure of the stable spline estimator with the atomic 
approach of section \ref{secANprior}, also introducing the concept of stable spline atoms.

\subsection{Stable spline kernels}

%Our aim is to derive a regularizer that encodes information on smoothness and exponential stability.
According to its stochastic interpretation, we have seen that every ReLS can be seen as 
a Bayesian estimator. In particular, each quadratic penalty can be obtained modeling the impulse response as a 
particular (zero-mean) Gaussian process. This implies that fixing the covariance (kernel) is equivalent to fixing the quadratic regularizer.\\ 
In system identification the kernel should include 
information on smoothness and exponential stability of the impulse
response $g$. One choice suggested in the literature is the so called first-order
Stable Spline kernel \cite{PillACC2010,ChenOL12}.
\color{black1} Letting $\E$ denote expectation, \color{black} for $t=1,2,\ldots$ and $s=1,2,\ldots$ it is defined by
\begin{equation} \label{SS1}
K(s,t)=\E [g_s g_t] \propto \alpha^{\max(s,t)}, \  0 \leq \alpha <1,%,   \quad (x_1,x_2) \in \mathbb{R}^+ \times \mathbb{R}^+  
\end{equation}
while it is null elsewhere. 
The second-order version was proposed in \cite{SS2010}. It is given by
\begin{equation}
\label{SS2}  \frac{\alpha^{s+t}
\alpha^{\max(s,t)}}{2}-\frac{\alpha^{3\max(s,t)}}{6},  \  0 \leq \alpha <1
\end{equation}
and leads to smoother impulse response realizations.
Notice that both  kernels
are parametrized by $\alpha$ which is interpreted as an unknown hyperparameter.\\

The left panel of Fig. \ref{Fig3} reports the standard deviations of a random vector $g$ % and $m=99$
whose covariance is the sampled version of (\ref{SS1}) with $\alpha$ set to \color{black1}  $0.8$ and $0.9$ respectively  (dashdot lines)\color{black} . 
Differently from the bathtub prior induced by the Hankel nuclear norm (solid line), the stable spline kernel  
describes system dynamics which go exponentially to zero. 
Further, \color{black1}  the \color{black}  hyperparameter $\alpha$ 
enhances model flexibility since it permits to tune the decay rate.
\color{black1}
Also, while nonstationary white noise underlies the Hankel prior (see the right panel of Fig. \ref{Fig3}), 
(\ref{SS1}) introduces correlation among the impulse response coefficients, %[REVIEW]
hence including information on impulse response smoothness.
This is important to have good MSE properties
as discussed in Remark \ref{rem:hanker:MSE}. With the same remark in mind,
note also that, differently from the Hankel nuclear norm
case, the stable spline prior shape is independent of the selected FIR order, thus making its effect on the estimation process more transparent. 
If $m$ is increased e.g. to $2m$, 
the statistics of the first $m$ impulse response coefficients remain the same.
\color{black}

%beta expected decay rate, dirlo dopo in confronto con Hankel nuclear nor.

%The covariance obtained above is exactly proportional to the Stable Spline
%kernel of order 1 introduced in \cite{PillACC2010} which thus enjoys
%favorable MaxEnt properties. Further, its discrete-time version 
%corresponds to the TC kernel introduced in  \cite{ChenOL12}  which, interestingly,
%was derived invoking a totally
%different deterministic argument. Notice that the kernel
%is parametrized by $\beta$ that is interpreted as an unknown hyperparameter.\\ 
%%Recalling (\ref{Wmkernel}), simple computations show that the RHS of (\ref{SS1}) 
%%is equal to $W(e^{-\beta s}, e^{-\beta t})$ for $m=1$.
%It is interesting to see that, as $\lambda_0$ goes to infinity, the process $h$ 
%in Proposition \ref{MaxEntprop} escapes from the class of differentiable processes.
%In fact, it tends to a Brownian motion (Wiener process),
%i.e. an integrated white noise (see Appendix \ref{A2} for other insights). 
%This also suggests that smoother descriptions of the impulse response can
%be obtained modeling $h$ as the $m$-fold integration of white
%Gaussian noise, with $m>1$. For instance, when $m=2$, one obtains the second-order
%stable spline kernel introduced in \cite{SS2010},
%i.e.
%\begin{equation}
%\label{SS2} \frac{e^{-\beta (s+t)}
%e^{-\beta \max(s,t)}}{2}-\frac{e^{-3\beta \max(s,t)}}{6}.
%\end{equation}

\subsection{Stable spline atoms and regularizer}

\color{black1}
Let $g$ be a discrete-time Gaussian stochastic process with covariance 
proportional to the stable spline kernel (\ref{SS1}).
Then, the following proposition characterizes 
the architecture of the minimum variance estimator of $g$ given the measurements (\ref{MeasMod}). 
%The proof (which is omitted) relies on the connection between 
%Bayes estimation of Gaussian processes (see Sections 1.4 and 1.5 in \cite{Wahba1990}) 
%and reproducing kernel Hilbert space (RKHS) theory. In particular, the use of
%the RKHS representation in Theorem 4 on pag. 37 of \cite{Cucker01}
%allows us to link ReLS with atomic approaches. 
%The estimator (\ref{MV}) can be also obtained by 
%the representer theorem for system identification (Theorem 3 on pag. 671 of \cite{SurveyKBsysid}).

\begin{proposition}\label{PropAtom}
Let
$$
y_i =  (g \otimes u)_i  +e_i, \quad i=1,\ldots,N
$$
where $e_i$ are all mutually independent, Gaussian, with zero-mean and variance $\sigma^2$. 
Assume also that $g$ is a zero-mean Gaussian and causal process, independent of the noise, with covariance %admitting the expansion
\begin{equation}\label{KerExp}
\E [g_s g_t] = \lambda \alpha^{\max(s,t)}
\end{equation}
with $\lambda$ a positive scale factor. 
Then, one has %the minimum variance estimate of $g$ given the $y_i$ is
\begin{equation}\label{ExpansionSS}
 \alpha^{\max(s,t)} = \sum_{j=1}^\infty  \zeta_j \rho_j(s) \rho_j(t)
\end{equation}
where
\begin{equation}\label{Exp1}
\rho_j(t) = \sqrt{2} \sin\left( \frac{\alpha^{t}}{\sqrt{\zeta_j}}  \right), \quad  \zeta_j = \frac{1}{(j\pi - \pi/2)^2}. 
\end{equation}
and
\begin{equation}\label{MVatom}
\E\left[ g_t | Y \right] = \sum_{j=1}^{\infty} \hat{a}_j \rho_j(t) 
\end{equation}
where
\begin{equation}\label{MVatom2}
\hat{a} =\arg\min_a \ \sum_{i=1}^N \left(y_i - \sum_{j=1}^\infty h_{ij} a_j  \right)^2  + \gamma \sum_{j=1}^{\infty} \frac{a_j^2}{\zeta_j}
\end{equation}
with
$$
h_{ij} =(\rho_j \otimes u)_i, \quad \gamma = \frac{\sigma^2}{\lambda}. %[] sistema convoluzione
%h_{ij} =L_i[\rho_j], \quad \gamma = \frac{\sigma^2}{\lambda}. %[] sistema convoluzione
$$% [REVIEW]
Finally, the estimate admits also the closed-form expression
\begin{equation}\label{MV}
\E\left[ g_t | Y \right] = \sum_{i=1}^{N} \hat{c}_i (K(\cdot,t) \otimes u(\cdot))_i  
\end{equation}
where $\hat{c}_i $ are the components of the vector  $\hat c$%[REVIEW] explicitly written
\begin{equation}\label{Cexp}
\hat c:=\left( A+ \gamma I_N  \right)^{-1} Y
\end{equation}
with
\begin{equation}\label{A}
A_{ij} =  \sum_{t=1}^\infty   u(j-t) \left( \sum_{k=1}^\infty u(i-k) K(t,k) \right)
%\left( (K \otimes u)_i  \otimes u \right)_j   %\int_{0}^{+\infty} u(t_i-s) \int_{0}^{+\infty} u(t_j - \tau )K(s,\tau) d\tau ds.  
\end{equation}
\begin{flushright}
$\blacksquare$
\end{flushright}
\end{proposition}
\color{black}

%L_i\left[e^{-\beta \max(\cdot,t)}\right]  

%When the covariance of $g$ is proportional to the stable spline kernel (\ref{SS1}),
%(\ref{MVatom2}) becomes a reformulation of the stable spline estimator.
%Furthermore, the expansion
%(\ref{KerExp}) is available in closed-form\footnote{The expansion of  (\ref{SS2}) is also available in closed form but
%is more complicated, see \cite{SS2010} for details.}:  
%from \cite{PillACC2010} one has 
%\begin{equation}%\label{rho1}
%e^{-\beta \max(s,t)} = \sum_{j=1}^\infty  \zeta_j \rho_j(s) \rho_j(t)
%\end{equation}
%where
%\begin{equation}\label{Exp1}
%\rho_j(t) = \sqrt{2} \sin\left( \frac{e^{-\beta t}}{\sqrt{\zeta_j}}  \right), \quad  \zeta_j = \frac{1}{(j\pi - \pi/2)^2}. 
%\end{equation}

The result above leads to the following comments: 
\begin{itemize}
\item when the stable spline prior (\ref{SS1}) is used, according to (\ref{MVatom}), the impulse response estimate is 
searched in the subspace spanned by the functions $\rho_j$ given by (\ref{Exp1}).
It is then natural to call 
\color{black1}
\begin{equation*} % \small
\mathcal{A}_{\alpha} = \left\{  \sin\left( \frac{\pi \alpha^{t}}{2} \right),  \sin\left( \frac{3 \pi \alpha^{t}}{ 2}\right),\sin\left( \frac{5 \pi \alpha^{t}}{ 2}\right), .. \right\}
\end{equation*}
\color{black}
the (first-order) {\it{stable spline atomic set}}.  
Such a set is parametrized by the positive scalar $\alpha$ which measures the distance 
from instability.
Note also that the {\it{stable spline atoms}} $\rho_j$ are ordered in such a way that their energy content 
at high frequencies increases as $j$ augments and that their structure is much different w.r.t. that adopted in \cite{Chand2012} and reported
in (\ref{Atset});
%Note also that their number is infinite and that there is no need
%to select the number of basis functions entering the impulse response description;
%we did see  the nuclear norm, in Bayesian terms provides a poor description of $g$ features and that 
\item atomic norms are typically designed in such a way as to assign the same penalty to each expansion coefficient $a_j$. 
Instead, according to (\ref{MVatom2},\ref{Exp1}), given $g=\sum_{j=1}^{\infty} a_j \rho_j$, 
the (first-order) stable spline estimator 
uses the {\it{stable spline regularizer}}
\begin{equation}\label{MaxReg}
J(g) = \sum_{j=1}^{\infty} \frac{a_j^2}{\zeta_j}
\end{equation}
where $\zeta_j$ are weights decaying to zero.  
The expansion coefficients $a_j$ are thus constrained to decay to zero at a rate which guarantees both impulse response continuity 
and system stability.
%\footnote{More specifically, the regularizer (\ref{MaxReg}) is forcing the estimate to belong to the reproducing kernel Hilbert space (RKHS) 
%of stable impulse responses associated with $K$. In fact, 
%if (\ref{KerExp}) holds, the Moore-Aronszajn theorem not only establishes a one-to-one correspondence between RKHS and positive semidefinite kernels \cite{Aronszajn50}, but also a precise characterization of this space: 
%every function in the RKHS is a combination either of 
%a finite number of $\rho_j$ or of an infinite number provided that the expansion coefficients $a_j$ satisfy
%$\sum_{j=1}^{\infty} a_j^2 / \zeta_j< \infty $.}. 
%\color{black1}
%This can be also 
%well appreciated in continuous-time where one also has (see Appendix \ref{A2} for the derivation)%[REVIEW]
%\begin{equation}\label{JSS}
%J(g) = \int_0^{\infty} \dot{g}^2(t) \frac{e^{\beta t}}{\beta} dt.
%\end{equation}
%\color{black}
%two key features neglected also by the Hankel nuclear norm, as already illustrated in Fig. \ref{Fig3}. 
%From (\ref{MVatom2}), one can also see that the regularizer is scaled by $\gamma$ which thus establishes the amount of regularization to be introduced. 
Hence, penalizing high-frequency components, also by adjusting $\gamma$, the stable spline regularizer may privilege more parsimonious 
models\footnote{Note that, when using the approach proposed in \cite{Shah2012}, the atoms reported in (\ref{Atomic2}) contain 
the term $1-|w|^2$ which implicitly penalises basis functions
close to the unit circle. However, no penalty is assigned to high-frequency impulse response components.}, possibly leading to estimators which better trade-off bias and variance;%accounting for the fact that, in real systems, the frequency response $G(j\omega)$ is expected to go to zero as 
%$\omega \rightarrow 0$;
%\item the architecture of the estimator (\ref{MVatom2}), equipped with the MaxEnt atomic set and the MaxEnt regularizer, depend on parameters $(\gamma,\beta)$ which admit a clear interpretation in frequency. In particular, $\gamma$ establishes the amount of penalty to be assigned to the high-frequency components of $g$ while $\beta$ measures the distance from instability, i.e. the distance of the atomic set from the imaginary axis;% (or from the unit circle in discrete-time);
\item the first step in the design of atomic-norm based estimators is the selection of the 
atomic set whose convex hull then defines the regularizer. 
In our stochastic framework, both of these steps are condensed 
in the choice of the covariance (kernel).
Indeed, the kernel encodes both the atoms $\rho_j$ and the regularizer,
defined by $\zeta_j$. % (we will elaborate further on this in the next subsection). 
This subsumes modeling and computational advantages: one needs neither % it is required neither  
to choose the number of atoms to be used (the kernel includes an infinite number of basis functions  $\rho_j$
so that no truncation affects the impulse response representation)  
nor store any basis function in memory. In fact, the estimate (\ref{MV})  
can be computed using the kernel just inverting a matrix, as shown in (\ref{Cexp}). 
This feature is related to what is called the {\it{kernel trick}} in the machine learning literature \cite{Scholkopf01b}.
%This has also computational advantages w.r.t. 
%Hankel. In fact, implementation of ReLS equipped with the Hankel nuclear norm 
%requires the solution of a convex optimization problem. This can significantly increase computational complexity since, to tune the regularization parameter e.g. via cross validation, typically many different estimates must be computed.
\end{itemize}

\section{Integral versions of stable spline kernels}
\label{Sec6}

\color{black1}
When the stable spline kernel is used, we have seen from
(\ref{Exp1}) that the atoms $\rho_j$ depend on a single parameter
$\alpha$ which establishes the decay rate of the eigenfunctions. 
Following also \cite{ChenTAC2014}, a way to further enrich the hypothesis space, 
making it more flexible, is to exploit different values of $\alpha$. 
However, differently from \cite{ChenTAC2014},  
our aim is the synthesis of a new kernel (in closed form) able to ``contain"
an infinite number of different decay rates, but which remains function only of a 
finite number of hyperparameters. 
In our Bayesian framework, we can obtain this by modeling the impulse response $g$
as the sum of i.i.d.  stochastic processes
whose stable spline kernels differ in the value of $\alpha$.
In particular, let us introduce two hyperparameters: $\alpha_m$, related to the fastest system pole,
and $\alpha_M$, connected with the slowest dynamics.
Then, we can consider $p$ values $\alpha_i$ 
satisfying $\alpha_m \leq \alpha_1 < \alpha_2 < \ldots < \alpha_p \leq \alpha_M$ and equally spaced with step $\Delta_{\alpha}$, then 
building the kernel 
\begin{equation}\label{AvKer}
\Delta_{\alpha} \sum_{i=1}^p  \alpha_i^{\max(s,t)}
\end{equation}
which induces the richer atomic set
\begin{equation}\label{AtomicSet2}
\mathcal{A}_{[\alpha_m,\alpha_M]} = \left\{  \sin\left( \frac{\alpha_i^{t}}{\sqrt{\zeta_j}}  \right), \ j=1,2,\ldots \ \mbox{and}  \ i=1,2,\ldots,p \right\}
\end{equation}
where we still have $\zeta_j = \frac{1}{(j\pi - \pi/2)^2}.$
Letting $p \rightarrow \infty$, so that $\Delta_{\alpha} \rightarrow 0$ in (\ref{AvKer}), the sum becomes the integral
$\int_{\alpha_m}^{\alpha_M} \alpha^{\max(s,t)} d\alpha$
which leads to the new {\it{first-order integral stable spline kernel}} called iTC and reported in 
Table \ref{TabSS}. The same procedure can be repeated starting from (\ref{SS2}) obtaining
the {\it{second-order integral stable spline kernel}}, called iSS in 
Table \ref{TabSS}.\\
Finally, these two kernels iTC and iSS can be summed up, obtaining
the kernel dubbed iTS in Table \ref{TabSS}.
This kernel thus synthetizes 
an infinite number of atoms that not only have different
decay rates $\alpha \in [\alpha_m,\alpha_M]$ but also two different smoothness levels.
\color{black}

%\section{ReLS based on stable kernels and Monte Carlo results}
%\label{Sec6}

%\subsection{Stable kernels-based estimators in discrete-time}\label{SSest}

\begin{table}
%{\bf{Stable Kernels}}
\begin{eqnarray*}
\hline
%\phantom{|}
%\label{eq:TC}\nonumber
\textbf{TC} \ 
&& P_{kj}(\eta)=\lambda\alpha^{\max(k,j)}; \\ 
&& \lambda\geq0, \; 0\leq\alpha<1, \; \eta=[\lambda,\alpha] \\
&& \medskip \\
\textbf{SS} \
&& P_{kj}(\eta) =  \lambda  \left(  \frac{\alpha^{k+j+\max(k,j)}}{2}  -  \frac{\alpha^{3\max(k,j)}}{6} \right)\\
&& \lambda \geq 0, \ 0 \leq \alpha<1, \ \eta=[\lambda,\alpha]\\
% \label{eq:DC}\nonumber  
%&& \medskip \\
%\textbf{DC} \
%&& P_{kj}(\eta)=\lambda\alpha^{(k+j)/2}\rho^{|j-k|}; \; \\
%&& \lambda\geq 0,  \; 0\leq\alpha<1,|\rho|\leq1;\ \;  
%\eta=[\lambda,\alpha,\rho]  \\
&& \medskip \\
\textbf{iTC} \ 
&& P_{kj}(\eta)=\lambda\frac{\alpha_M^{\max(k,j)+1} - \alpha_m^{\max(k,j)+1}}{\max(k,j)+1}\\
&& \lambda\geq 0,  \; 0\leq\alpha_m\leq\alpha_M<1;\ \;  
\eta=[\lambda,\alpha_m,\alpha_M]  \\
&& \medskip \\
\textbf{iSS} \ 
&& P_{kj}(\eta)=\frac{\lambda \alpha_M^{k+j+\max(k,j)+1}}{2\left(k+j+\max(k,j)+1\right)}
- \frac{\lambda \alpha_M^{3\max(k,j)+1}}{18\max(k,j)+6} \\
&& - \frac{\lambda \alpha_m^{k+j+\max(k,j)+1}}{2\left(k+j+\max(k,j)+1\right)}
+ \frac{\lambda \alpha_m^{3\max(k,j)+1}}{18\max(k,j)+6} \\
&& \lambda\geq 0,  \; 0\leq\alpha_m\leq\alpha_M<1;\ \;  
\eta=[\lambda,\alpha_m,\alpha_M]  \\
&& \medskip \\
\textbf{iTS} \ 
&& P_{kj}(\eta)= \lambda \frac{\alpha_M^{\max(k,j)+1 } - \alpha_m^{\max(k,j)+1}}{\max(k,j)+1}\\
&& + \frac{\lambda \alpha_M^{k+j+\max(k,j)+1}}{2\left(k+j+\max(k,j)+1\right)}
- \frac{\lambda \alpha_M^{3\max(k,j)+1}}{18\max(k,j)+6} \\
&& - \frac{\lambda \alpha_m^{k+j+\max(k,j)+1}}{2\left(k+j+\max(k,j)+1\right)}
+ \frac{\lambda \alpha_m^{3\max(k,j)+1}}{18\max(k,j)+6} \\
&& \lambda\geq 0,  \; 0\leq\alpha_m\leq\alpha_M<1;\ \;  
\eta=[\lambda,\alpha_m,\alpha_M]  \\
\hline
\phantom{|}
\end{eqnarray*}
\caption{{\bf{List of regularization matrices $P$ built using Stable Kernels}}}
\label{TabSS}
\end{table} 

%The first two matrices (TC and SS) are obtained by sampling the first and second order
%stable spline kernel given by (\ref{SS1}) and (\ref{SS2}), respectively (this follows the terminology adopted in 
%\cite{ChenOL12} and \cite{SurveyKBsysid}). %The third one is the so called DC kernel introduced in \cite{ChenOL12}. 
%The last three regularization matrices are obtained after integrating 
%TC and SS w.r.t. $\alpha$ over the interval $[\alpha_m,\alpha_M]$,
%in the same spirit of (\ref{iSS1}), (\ref{iSS2}) and (\ref{iSS1+2}).

We can now introduce estimators based on the different regularization matrices $P$ 
which are function of the hyperparameter vector $\eta$ as reported in Table \ref{TabSS}.
%In particular, the $(k,j)$ entry of the regularization matrix $P$ can be defined in five
%different ways, according to the list of stable kernels reported below.
The estimators have the same structure as  
documented in \cite{SS2010} and \cite{ChenOL12}. In particular,
assuming a high order FIR, 
it is useful to rewrite the measurements model (\ref{MeasMod}) 
as 
\begin{equation}\label{FIRmod}
Y= \Phi g + E
\end{equation}
Here, as in section \ref{Sec4}, $g$ now denotes the $m$-dimensional vector 
whose components are the impulse response coefficients % of the impulse response $g$
while the regression matrix $\Phi$ is defined by the input samples.
Then, the noise variance $\sigma^2$ is estimated from the residuals
obtained by fitting $g$ via least squares. 
The hyperparameter vector 
is instead determined through marginal likelihood optimization \cite{MLECC2014}, i.e. % optimizing the objective
\begin{equation}\label{MLopt}
\hat{\eta} = \arg\max_{\eta} \ Y^T \Sigma_{\eta}^{-1}  Y + \log \det \left( \Sigma_{\eta}  \right)
\end{equation}
where $\Sigma_{\eta}  = \Phi P \Phi^T + \sigma^2 I_n.$
Finally, the impulse response estimate is the solution of
$$
\arg \min_{g}   \|Y-\Phi g \|^2 + g^T P^{-1} g
$$
\color{black1} (note that $J(g)$ has become $g^T P^{-1} g$ above), \color{black} which is given by
\begin{equation}\label{SSforMC}
\hat{g} = P \Phi^T \Sigma_{\eta} ^{-1} Y 
\end{equation}
with $\eta$ set to its estimate $\hat{\eta}$. %their estimates.

\begin{figure*}
  \begin{center}
   \begin{tabular}{cccc}
\hspace{.1in}
 { \includegraphics[scale=0.42]{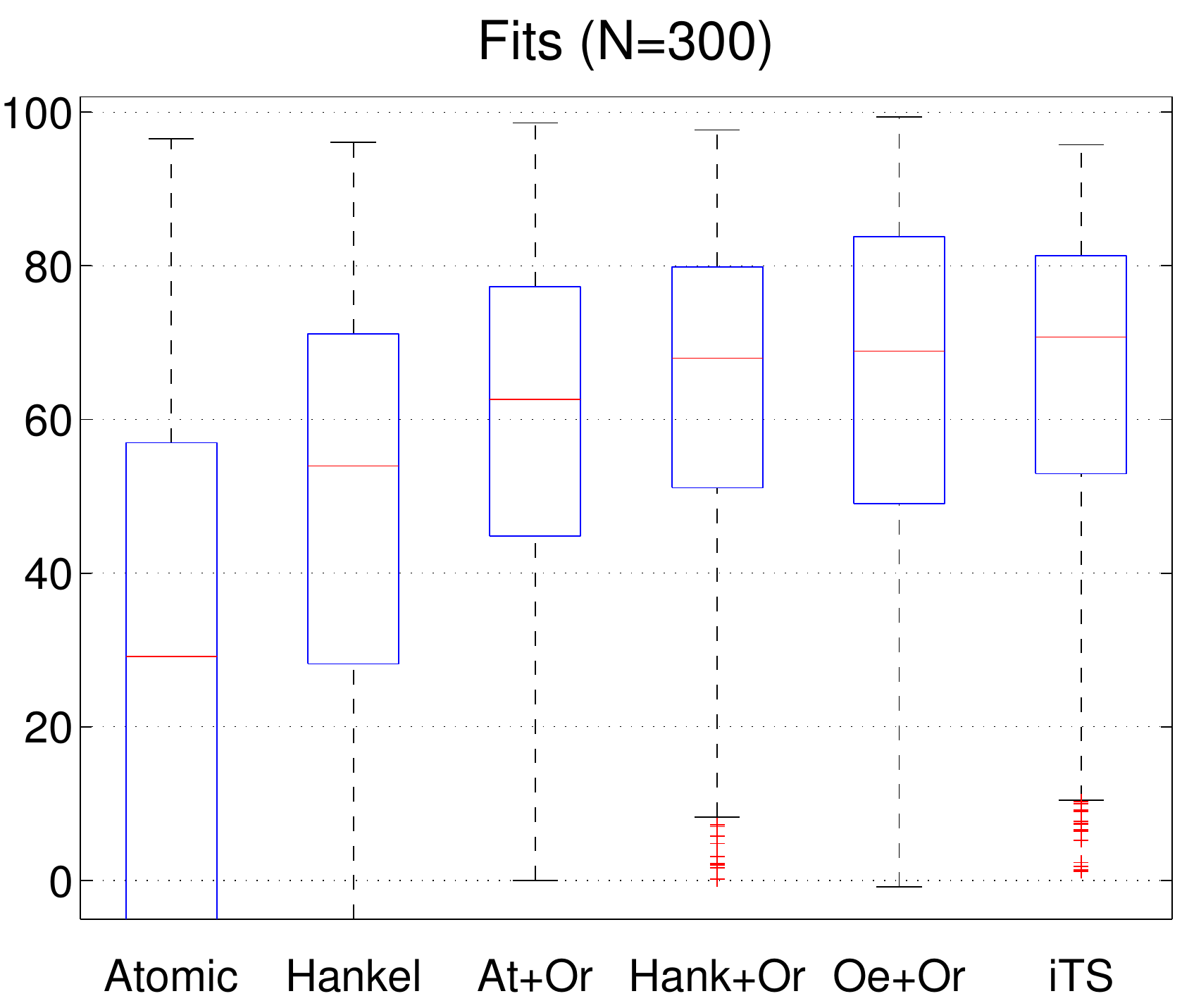}} 
\hspace{.1in}
 { \includegraphics[scale=0.42]{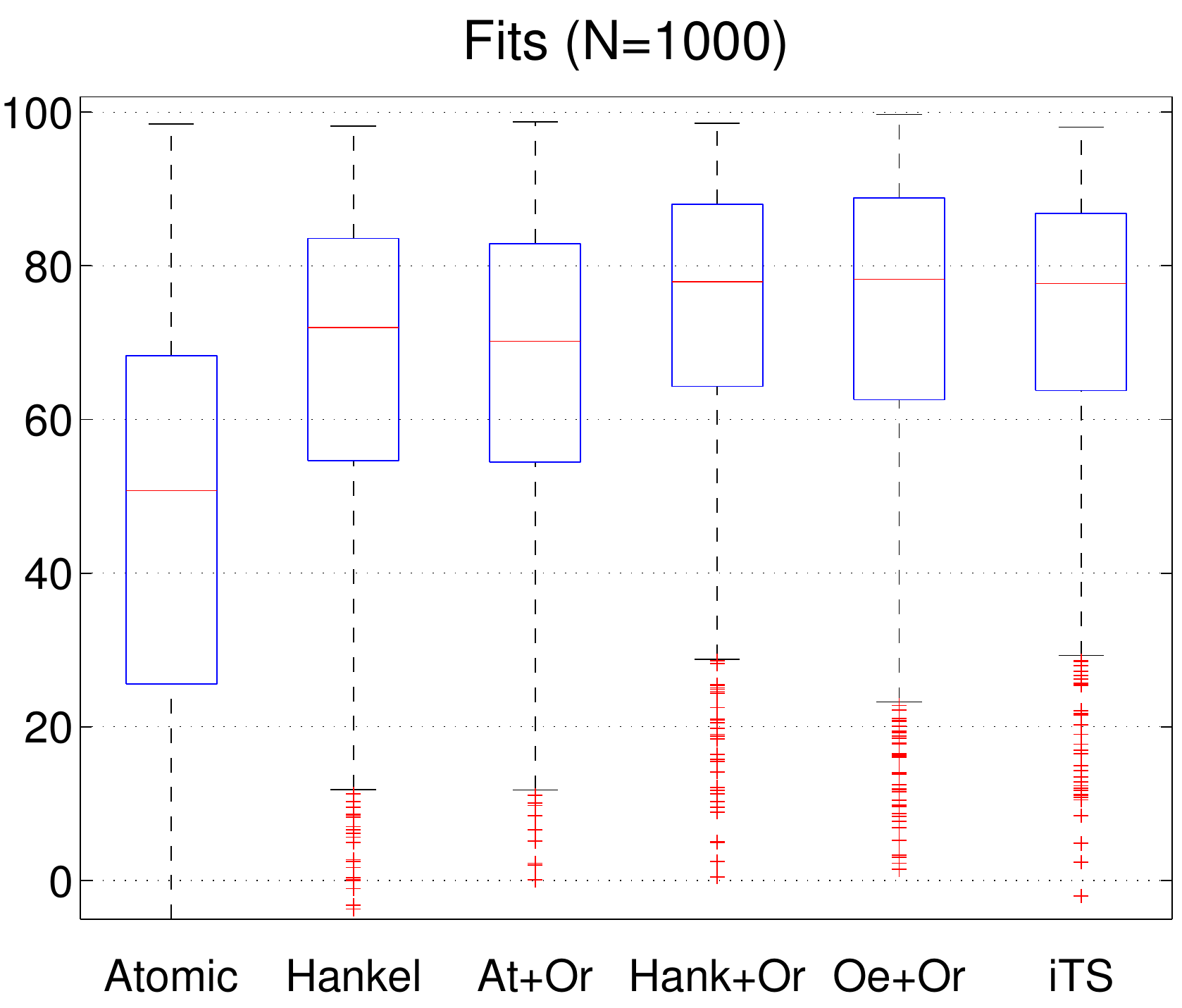}} 
    \end{tabular}
 \caption{{\bf{Monte Carlo experiment:}} boxplots of the fits using $300$ (left panel) or $1000$ (right panel) 
 input-output samples. Recall that At+Or, Hank+Or and Oe+Or are not implementable in practice
 since they exploit the knowledge of the true impulse response to tune model complexity.} 
    \label{Fig1}
     \end{center}
\end{figure*}

\begin{table*}
\begin{center}
\begin{tabular}{ccccccc}\hline
            & \textit{Atomic} & \textit{Hankel} & \textit{At+Or} & \textit{Hank+Or} & \textit{Oe+Or} & \textit{iTS} \\\hline
$N=300$   & 4.5 &  31.9  & 59.5 & 63.7 & 64.8 & 65.1\\
$N=1000$   & 39.2 & 63.3 & 67.1 & 73.7 & 72.6 & 72.9\\
 \hline
\phantom{|}
\end{tabular}
\end{center}
\caption{{\bf{Monte Carlo experiment:}}
average fit achieved by PEM equipped with oracle (Oe+Or), 
ReLS based on Atomic norms (Atomic and At+Or), Hankel nuclear norms (Hankel and Hank+Or) 
and the new estimator based on the stable kernel iTS, using $300$ or $1000$ input-output samples.  
Recall that At+Or, Hank+Or and Oe+Or are not implementable in practice
 since they exploit the knowledge of the true impulse response to tune model complexity. }
\label{Tab1}
\end{table*} 

\begin{figure*}
  \begin{center}
   \begin{tabular}{cccc}
\hspace{.1in}
 { \includegraphics[scale=0.42]{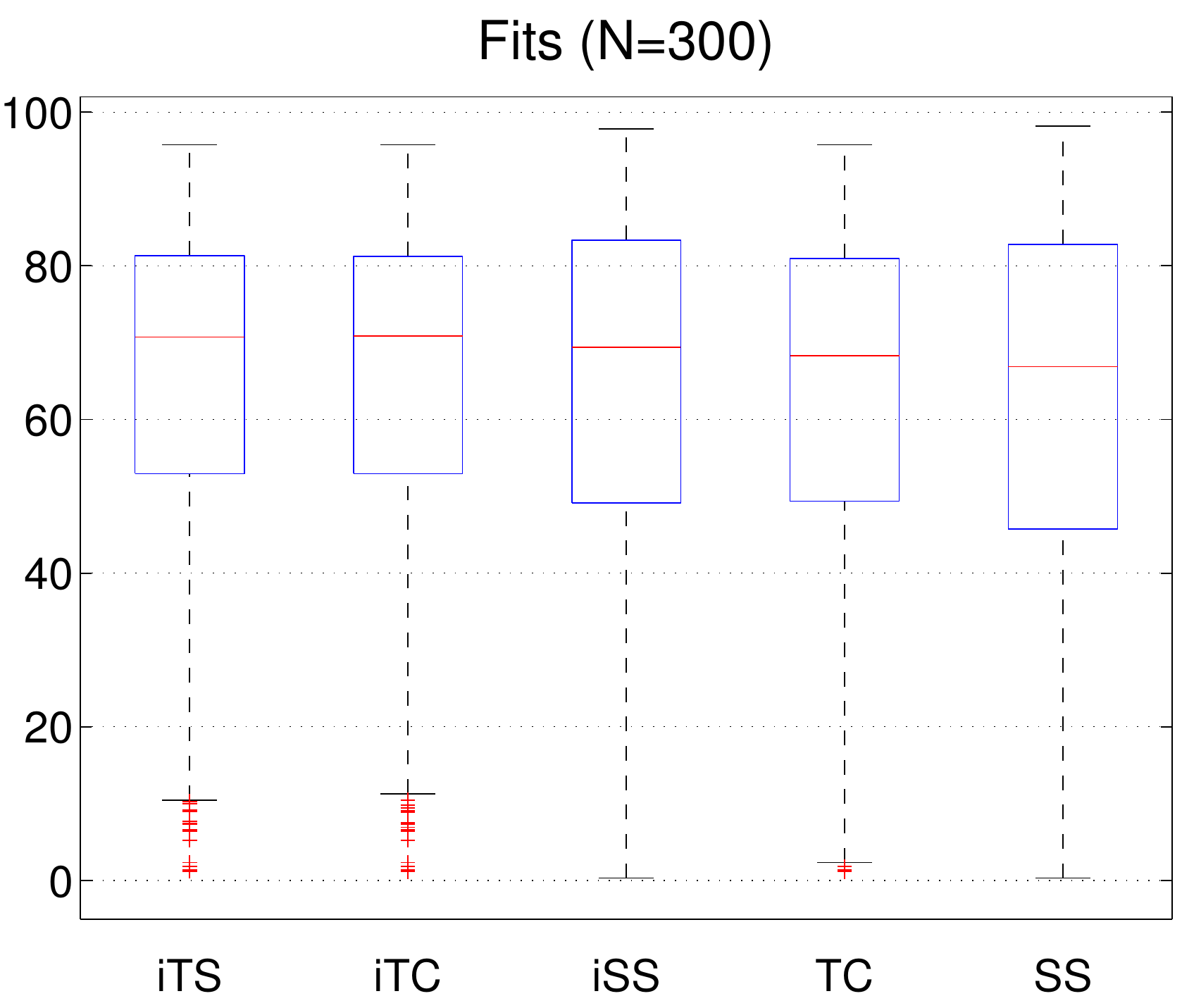}} 
 \hspace{.1in}
 { \includegraphics[scale=0.42]{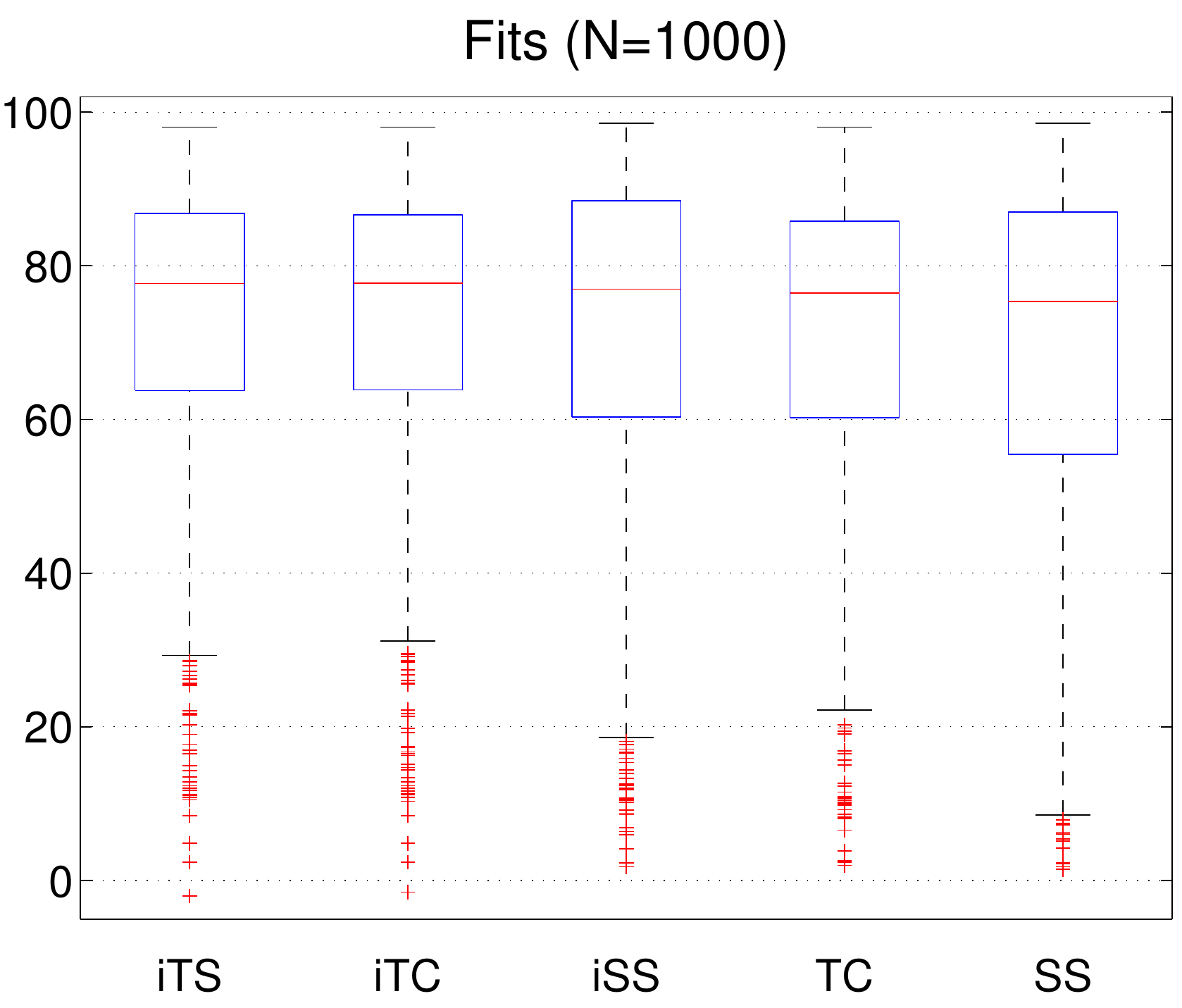}}
    \end{tabular}
 \caption{{\bf{Monte Carlo experiment:}} boxplots of the fits returned by ReLS equipped with different stable 
 kernels using $300$ (left panel) or $1000$ (right panel) 
 input-output samples. All the estimators are implementable in practice.} 
    \label{Fig2}
     \end{center}
\end{figure*}

\begin{table*} 
\begin{center} 
\begin{tabular}{cccccc}\hline 
            & \textit{iTS} & \textit{iTC} & \textit{iSS} & \textit{TC} & \textit{SS} \\\hline
$N=300$   & 65.1 &  64.8  & 64.4 &  63.2 &   62.3   \\
$N=1000$   & 72.9 & 72.7 &   71.6  & 70.6  &  69.2  \\
 \hline
\phantom{|}
\end{tabular}
\end{center}
\caption{{\bf{Monte Carlo experiment:}}
average fit returned by ReLS equipped with different stable 
 kernels using $300$ or $1000$ input-output samples. All the estimators are implementable in practice.}
\label{Tab2}
\end{table*} 

\begin{figure*}
  \begin{center}
   \begin{tabular}{cccc}
\hspace{.1in}
 { \includegraphics[scale=0.38]{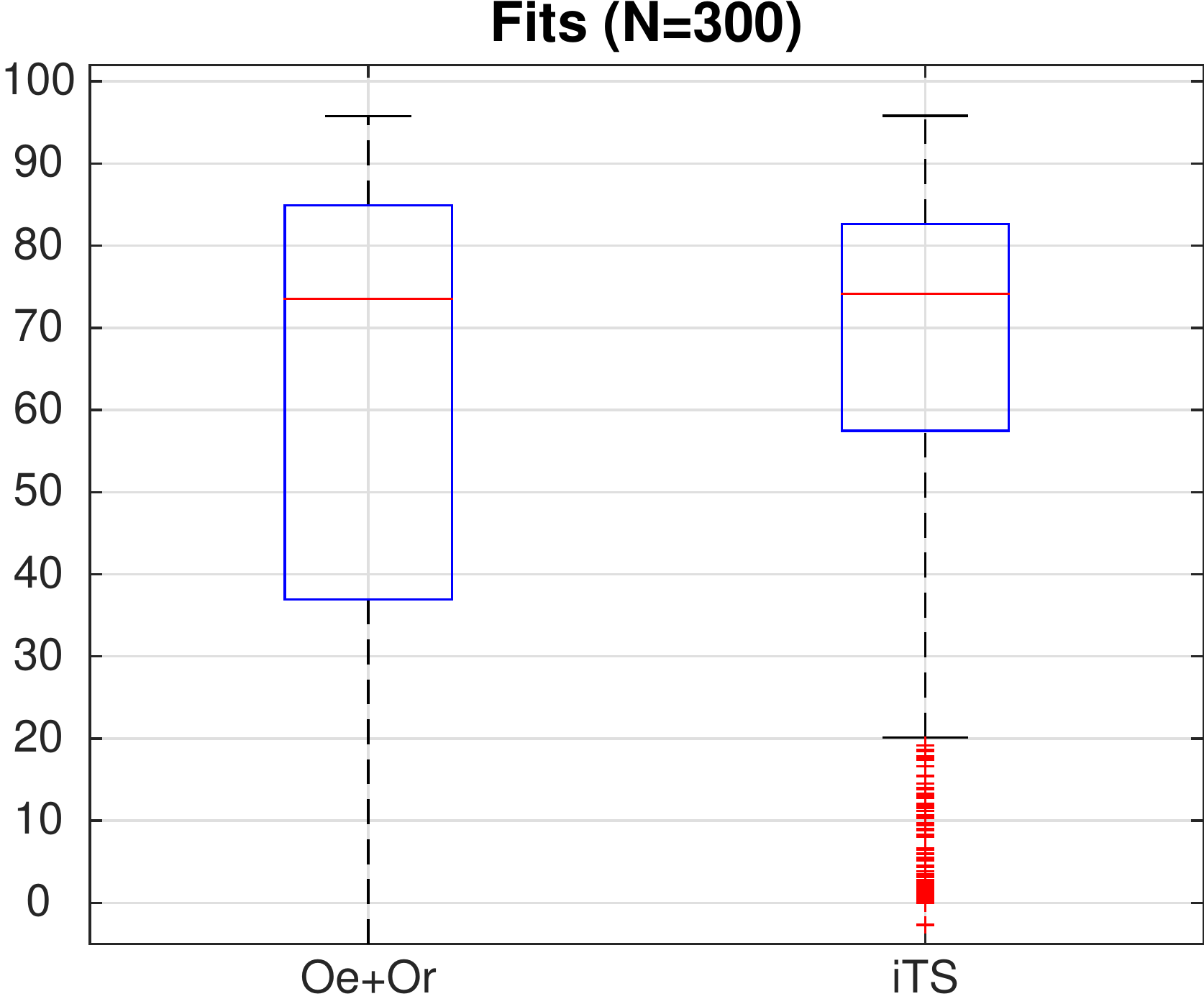}} 
 \hspace{.1in}
 { \includegraphics[scale=0.38]{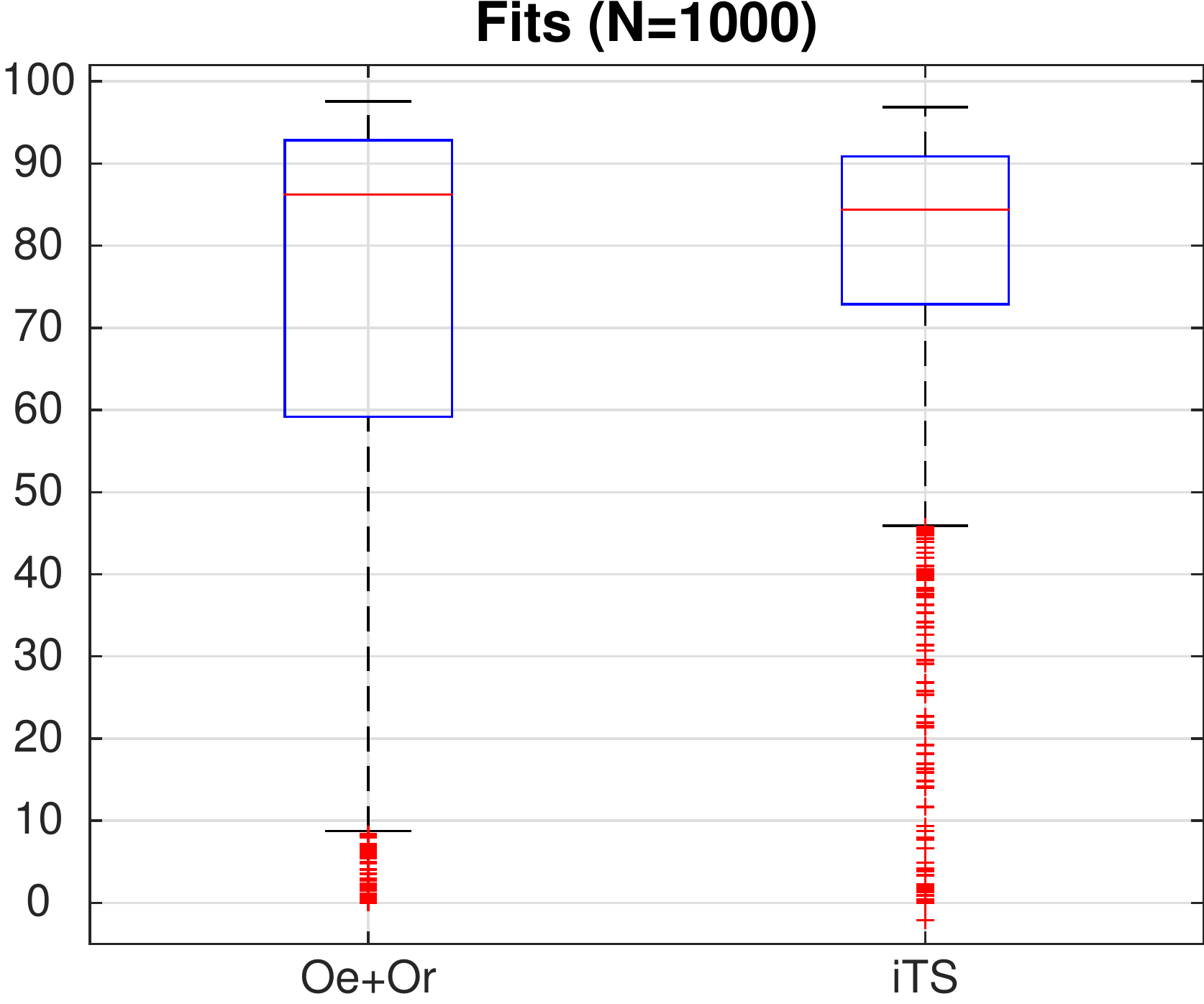}}
    \end{tabular}
 \caption{{\bf{Monte Carlo experiment using a different system generator:}} boxplots of the fits 
  achieved by Oe+Or and iTS using $300$ (left panel) or $1000$ (right panel) 
 input-output samples.} 
    \label{FigNewGen}
     \end{center}
\end{figure*}

%\begin{table*}
%\begin{center}
%\begin{tabular}{ccccccc}\hline
%            & \textit{iTS} & \textit{iTC} & \textit{iSS} & \textit{TC} & \textit{SS} & \textit{DC} \\\hline
%$N=300$   & 65.5 &  64.7  & 65.4 &  63.2 &   62.3   & 62.1\\
%$N=1000$   & 72.9 & 72.6 &   72.9  & 70.6  &  69.2 &   69.9\\
% \hline
%\phantom{|}
%\end{tabular}
%\end{center}
%\caption{{\bf{Monte Carlo experiment:}}
%average fit achieved by PEM equipped with oracle (Oe+Or), 
%ReLS based on Atomic norms (Atomic and At+Or), Hankel nuclear norms (Hankel and Hank+Or) 
%and the new estimator based on the stable kernel iTS.  Recall that At+Or, Hank+Or and Oe+Or are not implementable in practice
% since they exploit the knowledge of the true impulse response to tune model complexity. }
%\label{Tab2b}
%\end{table*} 

\section{Simulated data: Monte Carlo studies} 
\label{Sec7}

\subsection{Data sets and performance index} 

We compare different estimators for discrete-time system identification.
To this aim, we resort to two Monte Carlo studies whose
implementation details are the same as described in Section 7.2 of \cite{SurveyKBsysid}. Here, we just recall that 
each Monte Carlo consists of 1000 runs. At each run, $g$ is defined by a different rational transfer function, given by
the ratio of two polynomials of the same order (varying from 1 to 30), randomly obtained by a Matlab generator. 
The system, initially at rest, is fed with an input obtained by filtering    
a zero mean unit variance white Gaussian noise 
through a 2nd order rational transfer function which also varies from run to run.  
Output data are corrupted by a white Gaussian noise, with the SNR (ratio between the variance of the noiseless
output and the noise) randomly drawn at every run in the interval $[1,10]$.
The first and the second Monte Carlo study differ in the number $N$ of available input-output measurement, 
equal to 300 or 1000, respectively.\\
Given an estimator $\hat{g}$, its performance index 
is evaluated computing the fit measures % computed, at the $j$-th Monte Carlo run, as
\begin{equation}
\label{Fit}
\F_j = 100 \times \left(1 - \frac{ \| g_j - \hat{g}_j \|}{\|g_j\|}\right ), \quad j=1,\ldots,1000
\end{equation}
where $g_j$ and $\hat{g}_j$ are the true and the estimated impulse response at the $j$-th run.

\subsection{\color{black1}  Estimators compared via  the \color{black}  Monte Carlo study}\label{SevEst}

\paragraph*{Oe+Or} All the impulse response estimators introduced so far
depend on an unknown parameter vector, denoted by $\eta$, which controls model complexity. 
For instance, in all the regularized techniques $\eta$  contains at least
the regularization parameter $\gamma$.  
When using PEM, $\eta$ instead represents the order of different model structures.
For instance, consider the use of rational transfer functions 
for discrete-time system identification. Then, $\eta$ is
the degree of  the polynomials $B(z)$ and $A(z)$ composing 
the transfer function given, in the $z$-transfer domain, by
\begin{align}
  \label{eq:RTF}
  G(z)&=\frac{B(z)}{A(z)} 
\end{align} 
\color{black1} Hereby, $\hat{g}$ is said to be an oracle-based estimator if,  having access to the true impulse response $g$,
it determines  model complexity by maximizing the fit measure 
(\ref{Fit}) w.r.t. $\eta$. Note that such an impulse response estimator 
is never implementable in practice since  the true impulse response $g$ is not available.\color{black} 
This identification procedure is however useful since it provides a performance reference. 
In particular, $Oe+Or$ denotes the following PEM procedure. First,  
the Matlab function \texttt{Oe.m} is used to fit the model structures (\ref{eq:RTF}) to data
for $\eta=1,\ldots,30$ (the information that system was initially at rest is given to the estimator). 
Then, among the 30 impulse response estimates obtained,
$Oe+Or$ returns that maximizing (\ref{Fit}).  

\paragraph*{\{Hankel,Hank+Or\}} 
We have implemented two variants of the estimator (\ref{HankelEst})
based on the Hankel nuclear norm adopting a FIR model of order
$m=99$ with the size of the Hankel matrix $H(g)$ equal to $50 \times 50$.
In the first version, dubbed Hankel, the regularization parameter is 
estimated via cross validation. Data are divided into an identification and
validation data set of equal size. For every value of $\gamma$
in the grid defined by the MATLAB command $\texttt{\small logspace(-5,4,50)}$,
the solution (\ref{HankelEst}) is obtained from the identification data using the software CVX \cite{CVX1,CVX2}. 
The regularization parameter $\hat{\gamma}$ is the one leading to the best prediction on the validation set
according to a quadratic fit criterion. Finally, the impulse response estimate is achieved by solving (\ref{HankelEst}) 
using $\gamma=\hat{\gamma}$ and all the available data. The second version of the estimator
exploits an oracle which, at every run, selects the value of $\gamma$ in the set $\texttt{\small logspace(-5,4,50)}$
which maximizes the fit (\ref{Fit}).\\

\paragraph*{\{Atomic,At+Or\}} Two types of atomic estimators (\ref{Atomic2})
have been implemented. In both the variants, 
the atomic set is defined by the poles $w_k=\alpha e^{\sqrt{-1} \beta}$ 
and their complex conjugate $w_k^*$ where, using a MATLAB notation,
$\alpha$ and $\beta$ \color{black1}  take \color{black}  values on the two grids $\texttt{\small [0.02:0.02:0.98  \ 0.99 \ 0.999]}$
and $\texttt{\small [$0$ \ $\pi$/50:$\pi$/50:$\pi$]}$, respectively. 
Optimization (\ref{Atomic2}) is then performed constraining 
each couple of expansions coefficients $a_k$ 
related to complex conjugate poles to be \color{black1}  equal and real.\color{black} \footnote{We have also implemented the estimator (\ref{Atomic2}) equipped with a different atomic set
given by discrete-time Laguerre basis functions \cite{WahLag}. Results (not shown) are similar to those
described in the sequel obtained adopting (\ref{Atset}).}
The first estimator, dubbed Atomic, obtains at each Monte Carlo run
the value of the regularization parameter via 10-fold cross validation. 
It has been implemented exploiting the MATLAB software package
$\texttt{glmnet}$ \cite{glmnet}, providing the information on system
initial conditions and input delay. We have also used the MATLAB commands
$\texttt{\small options.lambda=logspace(-5,4,100)}$ to define the grid where the regularization parameter is searched
and $\texttt{\small options.intr=0}$ to specify that the relation between $g$ 
and the system output is linear (and not affine). The second estimator, dubbed At+Or, 
is implemented in the same way, except that 
the regularization parameter is selected by the oracle.

\paragraph*{\{TC,SS,iTC,iSS,iTS\}} These are the estimators 
(\ref{SSforMC}) equipped with the stable kernels
described in Section \ref{Sec6}, with the dimension of $g$ set to $m=100$ and 
the hyperparameter vector $\eta$ estimated via marginal likelihood optimization  
(\ref{MLopt}). Note that all of these estimators are implementable in practice

\subsection{Results}

\color{black1}  Boxplots of the fits \color{black}  achieved by Atomic, At+Or, Hankel, Hank+Or,
Oe+Or and iTS in the two Monte Carlo studies are displayed in 
the left and bottom panel of Fig. \ref{Fig1}. 
Table \ref{Tab1} also shows the average fits.\\
\color{black1} It is apparent \color{black}  that the fits of Hankel are significantly smaller than those
returned by Oe+Or. \color{black1}  The more so \color{black}  in the first experiment where the average fit of
Hankel is $31.9$ while that of Oe+Or is $64.8$. 
\color{black1} Instead,  the performance  of Hank+Or  ($\gamma$ is chosen by the oracle)
is virtually identical to that of Oe+Or. \color{black}
As for Atomic, its performance is not satisfactory, inferior than 
that of Hankel.  Only using the oracle-based estimator At+Or,
the performance becomes comparable with that of Oe+Or.
Instead, the integral prior iTS largely outperforms the 
other regularized system identification approaches 
proposed in the literature. Remarkably, its performance is very close or also superior to that of the
oracle-based approaches. For instance, in the two experiments 
iTS provides average fits equal to $65.1$ and $72.9$  
whereas Oe+Or return $64.8$ and $72.6$.\\ 
The beneficial effect of the unequal weighting on the expansion coefficients
is evident also reconsidering Fig. \ref{FigFix}: iTS (right panel, results with $\gamma$ estimated via marginal likelihood)
outperforms At+Or (left panel, results with $\gamma$ estimated by the oracle).\\
Finally, Fig. \ref{Fig2} and Table \ref{Tab2} permit to compare the performance of all the ReLS 
approaches equipped with the stable kernels. 
As a matter of fact, all the estimators perform well,
revealing the importance of informing the estimation process of system stability. 
Notice also that iTS provides the best results, pointing out 
benefits of integral versions of stable kernels.

\color{black1}
\subsection{Use of a different system generator}
The random generator  adopted  so far defines 
challenging systems, with poles often located at high frequencies.
However, a visual inspection reveals that the average  
number of significant Hankel singular values is rather small,
being around 10. A consequence is that the mean  order selected by Oe+Or is around 5. 
We have thus found of interest to repeat the experiment 
adopting a different ``higher-order"  generator. The rational transfer function order is now fixed to 30
and  the  poles are selected iterating the following procedure at every Monte Carlo run: With equal probability a real or a couple of complex conjugate poles is added to the denominator
until its order reaches  30.  In the case of a real pole,
it is randomly drawn from a uniform distribution on $[-0.95,0.95]$, while the absolute values and phases
of the complex conjugate pairs are independent random variables uniform on $[0,95]$ and $[0,\pi]$, respectively.
The zeros are then selected in the same way except that 
their absolute values are drawn in the interval $[-2,2]$.\\
Boxplots of the fit values are reported in Fig. \ref{FigNewGen}, restricting the comparison to Oe+Or and iTS.
Remarkably, the advantage of iTS over Oe+Or increases: its average fit is $65.3$ vs $59.3$
when $N=300$ and $76.7$ vs $73.3$ when $N=1000$. This can be explained 
considering that, on average,
Oe+Or now selects models of order $13$ and this may further undermine
optimal asymptotic properties of PEM under correct order specification. In addition, such an estimator has now to hinge on 
higher-dimensional non convex problems, possibly more prone to local minima.  
Under these circumstances, ReLS equipped with empirical Bayes
can be especially useful, see also \cite{MLECC2014,MLAut2015} for insights on  
marginal likelihood effectiveness in controlling model complexity.  
%Note also that iTS returns the estimated model optimizing a single marginal likelihood
%over a space with dimension always equal to 3.

\section{Real data: temperature prediction}\label{sec:RealData}

\begin{figure}
\begin{center} 
\begin{tabular}{cc}
\includegraphics[width=0.5\columnwidth,angle=-90]{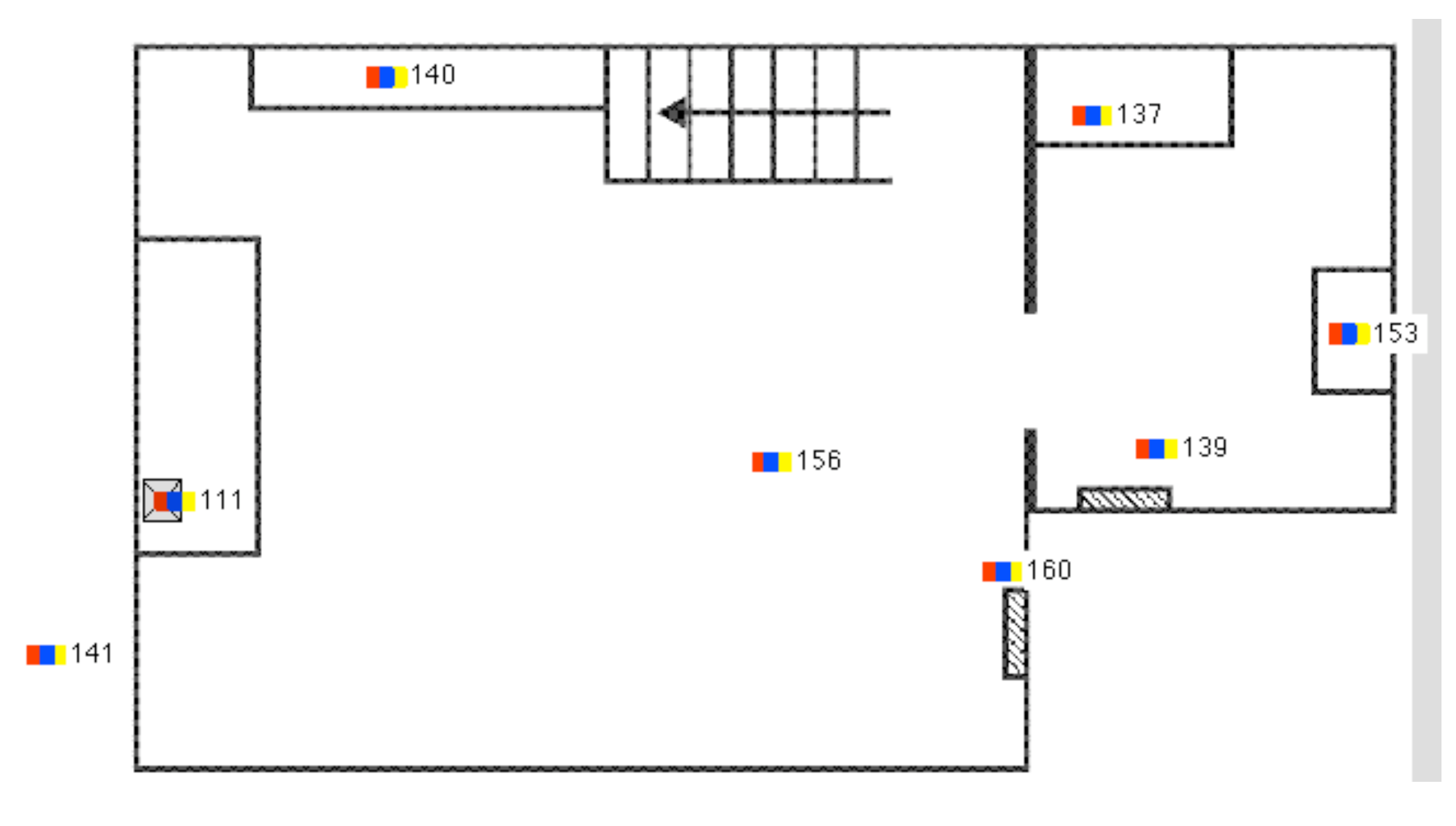} \\
\includegraphics[width=0.75\columnwidth,angle=0]{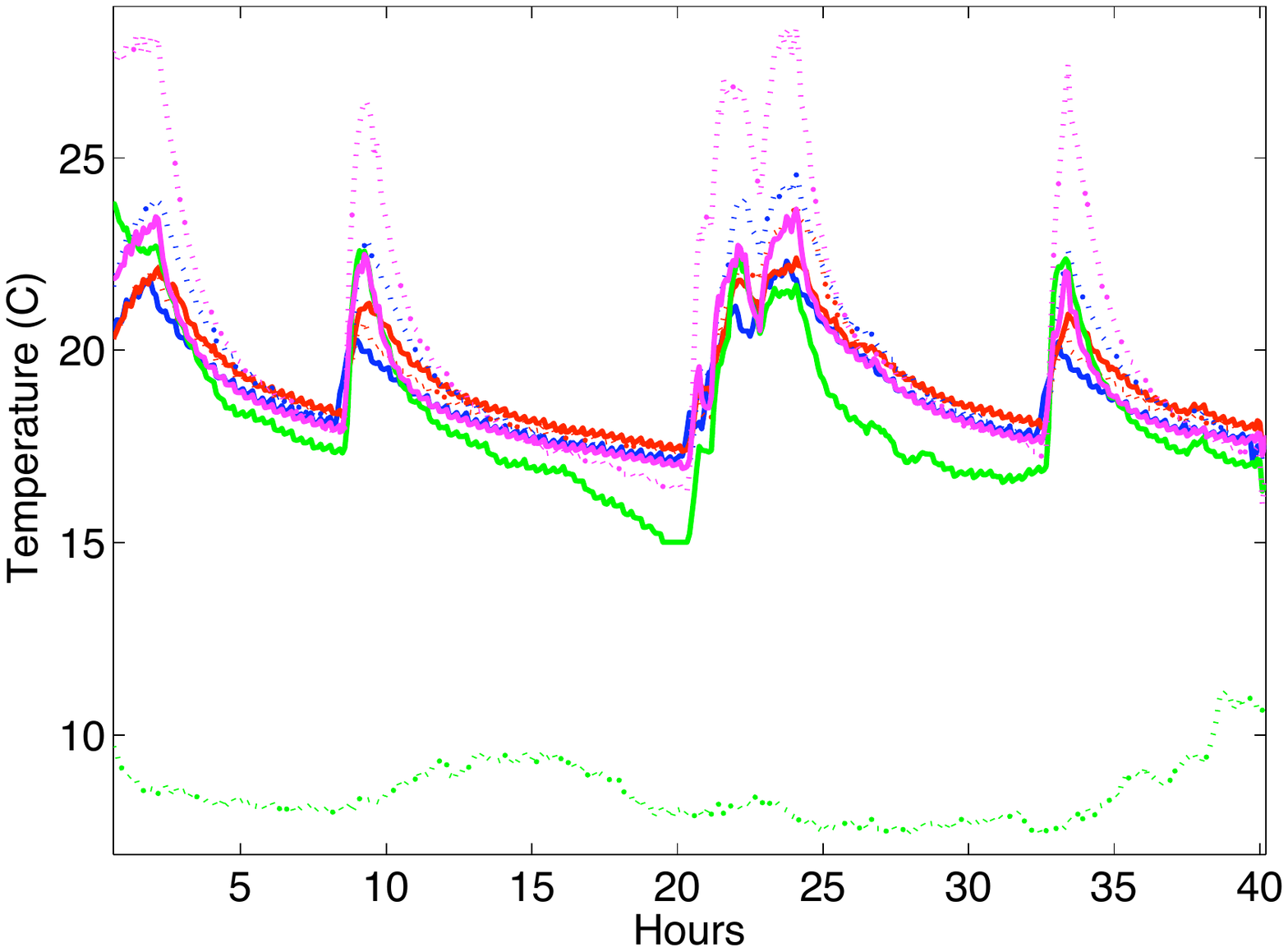}
\end{tabular}
\caption{Nodes location (top) and measured temperatures during the first 40 hours (bottom).} \label{signals}
\label{FigTemp}
\end{center}
\end{figure}

To test the algorithms on real data we have also considered thermodynamic modeling of 
buildings. We placed sensors in two rooms of a small two-floor residential building of about $80$ $\textrm{m}^2$ and $200$
$\textrm{m}^3$; the sensors have been placed only on one floor (approximately 40 $\textrm{m}^2$) and their location is approximately shown in  Fig. \ref{FigTemp} (top panel).
The larger room is the living room while the smaller is the kitchen.
The experimental data was collected through a 
wireless sensor network made of 8 \emph{Tmote-Sky} nodes
produced by Moteiv Inc, each of them is provided with a
temperature sensor. %, a humidity sensor, and a total solar radiation photoreceptor (visible + infrared). %(\cite{tmote-data-sheet}).
 The building was inhabited
during the measurement period, which lasted for 8 days starting from February 24th, 2011; samples
were taken every 5 minutes.
The heating systems was controlled by a thermostat; the reference temperature
was manually set every day depending upon occupancy and other needs.\\
 The location of the 8 sensors was as follows:

 \begin{itemize}
 \item Node $\#$1 (label 137 in Fig. \ref{FigTemp}) was above a cabinet (2.5 meters high).
  \item  Node $\#$2 (label 111) was above a sideboard, about $1.8$ meters high, close to thermoconvector.
 \item Node $\#$3 (label 139) was above a cabinet (2.5 meters high).
  \item  Node $\#$4 (label 140) was placed on a bookshelf (1.5 meters high).
 \item Node $\#$5 (label 141) was placed outside.
 \item Node $\#$6 (label 153) was placed above the stove  (2 meters high).
 \item Node $\#$7 (label 156) was placed in the middle of the room, hanging from the ceiling (about 2 meters high).
 \item Node $\#$8 (label 160) was placed above one radiator and was meant to provide
a proxy of water temperature in the heating systems.
\end{itemize}

This gives a total of $8$ temperature profiles.  A preliminary inspection of the measured signals, reported in
Fig. \ref{FigTemp} (bottom panel) reveals the high level of collinearity which is well-known to complicate the estimation process in System Identification \cite{Soderstrom,Ljung:99,BoxJenkins}.\\
We only consider
Multiple Input-Single Output (MISO) models,
with the temperature from the first node as output ($y_i$) and the other
7 temperatures as inputs ($u^j_i$, $j=1,..,7$).
We leave identification of a full Multiple Input-Multiple Output (MIMO) model for future investigation.  We split the available data into $2$ parts; the first $N_{id}=1000$ temperature samples 
are the identification data while the last $N_{test}=1500$ 
are used for test purposes.
The notation $y^{test}$ identifies the test data.
Note that $N_{id}=1000$, with $5$ minute sampling times,
corresponds to around $\simeq 80\;hours$;
this is a rather small time interval and, as such,
models based on these data cannot capture seasonal variations.
Consequently, in our experiments we assume a ``stationary'' environment and normalize
the data so as to have zero mean and unit variance before identification is performed.\\
We envision that model predictive based methodologies, (see \cite{ModelPredictiveBookCamacho} and the recent papers \cite{Borrelli2010}, \cite{Privara2011}, \cite{Dong2008}), may be effective for these applications and, as such, we evaluate our new estimators based on their ability to predict future data.
The predictive power of the model is measured for $k$-step-ahead prediction on \emph{test} data, as the fit:
\begin{equation}\label{FIT}
\F^k :=100 \times \left(1-{\frac{\sqrt{\sum_{i=k}^{N_{test}}(y^{test}_i-\hat y_{i|i-k})^2}}{\sqrt{\sum_{i=k}^{N_{test}} (y^{test}_i)^2}}}\right).
\end{equation}

\begin{figure*}
  \begin{center}
    \includegraphics[width=0.9\columnwidth,angle=0]{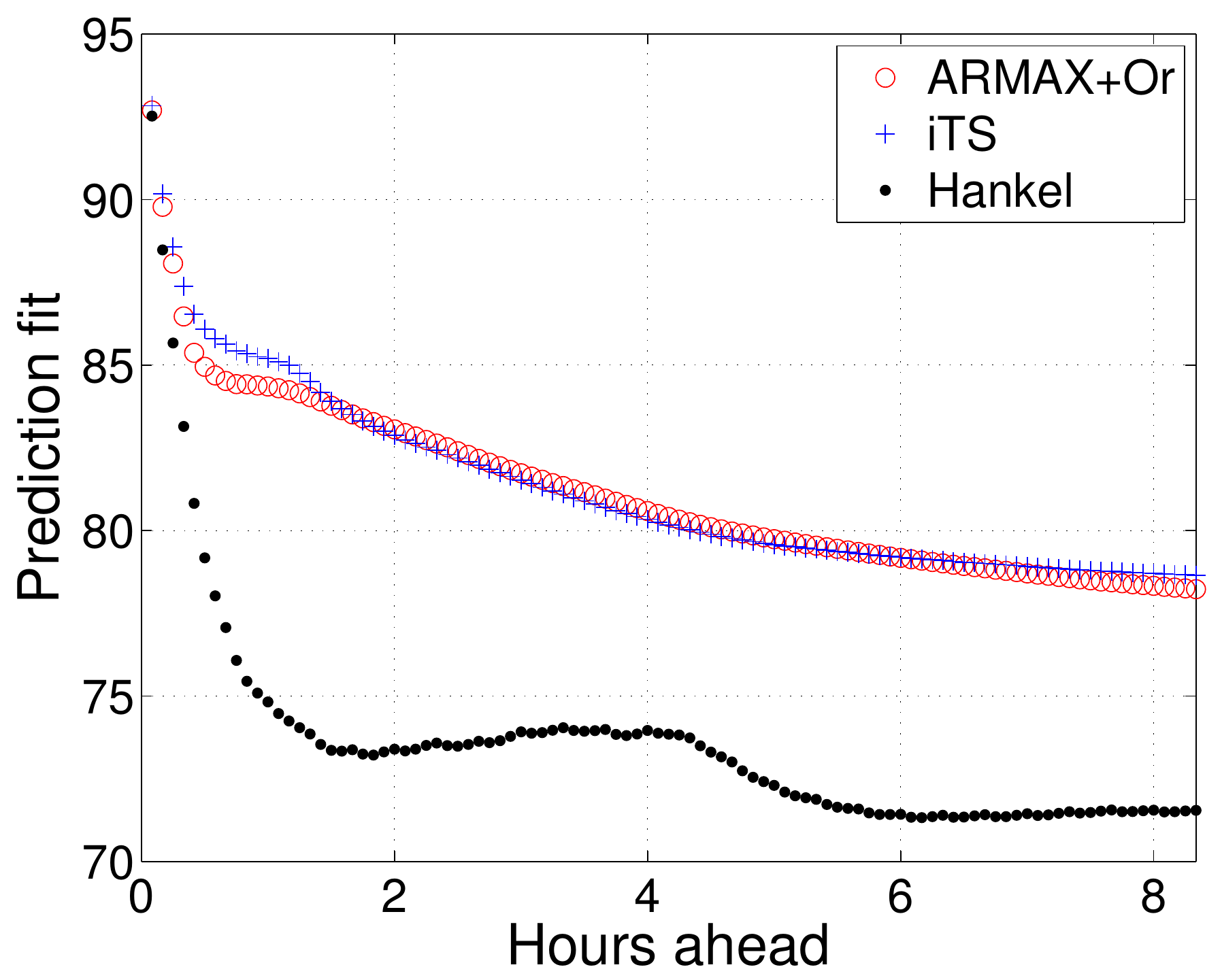}% \hfill
    \includegraphics[width=0.9\columnwidth,angle=0]{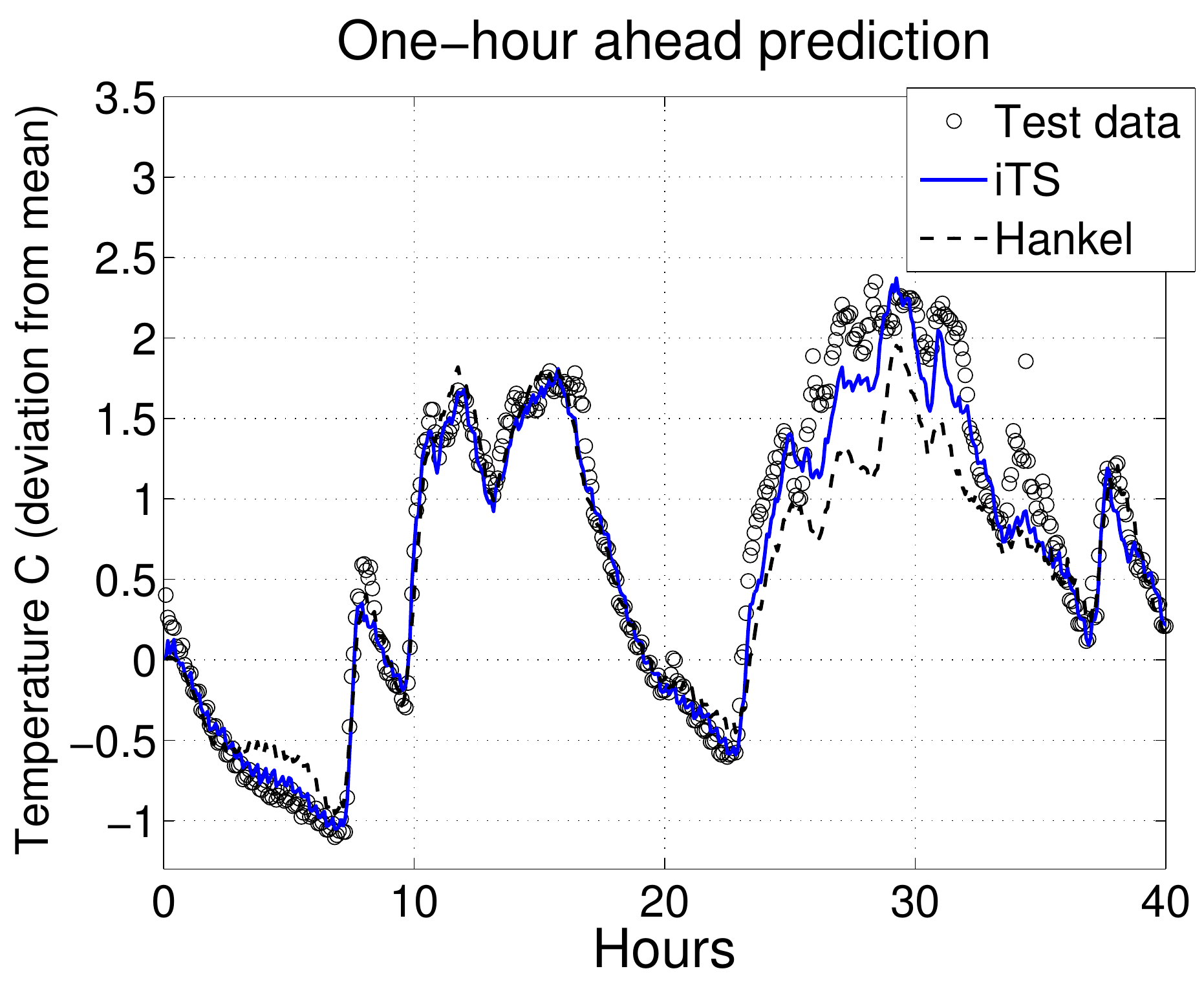}
    %\hfill
\caption{Test set prediction fit $\F^k$ as defined in (\ref{FIT})  (left) 
and 1-hour ahead test set prediction (right). Recall that Oe+Or is not implementable in practice
 since it exploits the knowledge of the test set to tune model complexity.}
 \label{FigTemp2}
  \end{center}
\end{figure*}

Identification has been performed using 
ARMAX+Or, Hankel and iTS. More specifically, ARMAX+Or 
exploits ARMAX models formed by polynomials of the same order. 
It has access to the test set and 
selects that model order which maximizes $\sum_{k=1}^{100} \F^k$.
In particular, it turns out that setting the order to 7 
provides the best average fit on an horizon of length around 8 hours.\\
As for the two regularized estimators, consider ARX models of the form
 $$
 y_i = (g^1 \otimes y)_i  + \sum_{j=1}^{7} (g^{j+1} \otimes u^j)_i + e_i,
$$
where the $\{g^j\}$ are the 8 unknown one-step ahead predictor impulse responses.
Then, both Hankel and iTS assume the form
\begin{equation}\label{HankelEst}
\arg \min_{g^i}  \ \sum_{i=1}^N \left( y_i - (g^1 \otimes y)_i  - \sum_{j=1}^{7} (g^{j+1} \otimes u^j)_i  \right)^2 + 
\gamma \sum_{j=1}^8   J(g^j),
%\arg \min_{g} \| Y - \Phi g \|^2 + \gamma \| H(g) \|_*
\end{equation}
differing only in the adopted $J$. Both these estimators are then implemented
in the same way described in the previous section.\\
The $3$ estimators are compared using as performance indexes  $\F^k$ defined in  \eqref{FIT}.
The results are reported in Fig. \ref{FigTemp2} (left panel): similarly to what happened using simulated data,
the performance of ARMAX+Or and iTS is similar and superior than that of Hankel.
Sample trajectories of one-hour-ahead test data prediction
obtained by iTS anzd Hankel are also visible in Fig. \ref{FigTemp2} (right panel). 
\color{black}

\section{Conclusions}

The results of this paper highlight 
the importance of Bayesian interpretation of ReLS for system identification.
In particular, the Bayesian framework offers transparent guidelines for 
selecting regularizers that capture crucial system features. 
Another use regards the assessment of existing regularizers: for instance,
the drawbacks of the Hankel nuclear norm have been linked 
to the adoption of a Bayesian prior which models the 
impulse response as a nonstationary white noise.\\
Similar drawbacks affect also other recent approaches that model the impulse response as the
superimposition of a large set of atoms, employing an atomic norm as regularizer. In the literature, it has been argued that any reasonable penalty function should be constant 
on the set of atoms \cite{Chand2012}. Actually,   \color{black1} this can result in penalty functions with poor capability of controlling model complexity, thus leading to estimators with large variance\color{black}.  Indeed, if the number of atoms to be included tends to  infinity,
all of them being e.g. of unit $\ell_1$ norm,  
any regularizer including smoothness and system stability can not assign the same penalty to the atoms.
Including smoothness and stability constraints in the estimation process, e.g. using stable spline kernels, 
instead leads to a kind of regularizer 
whose weights are non uniform. Advantages are confirmed by the
simulation studies: in the Monte Carlo experiments the family of
regularizers induced by stable kernels, including the three novel covariances  iTC, iSS and iTS, 
performs systematically better than other nuclear and atomic techniques.\\
\color{black1}
In conclusion, we also stress that, even if drawbacks 
of the Hankel norm have been here illustrated, this does not mean that
such  regularizer can not be useful for system identification.
For example, the MIMO case has not been well investigated yet and  there can be also 
cases where large magnitude corruptions
or un-modelled dynamics can be well described 
by Hankel or weighted Hankel norms \cite{Mohan2010,sadigh-ifac14}. An interesting perspective is also the
design of estimators that combine stable kernels and atomic norms.
Preliminary work on this can be found in  \cite{SSvsNN2013,ChiusoNN2014}.
\color{black}

%In the paper, we considered three types of stable kernels, namely SS/TC/DC, already
%available in the literature. They enjoy meaningful interpretation in terms of 
%estimation of Gaussian processes and regularization
%in RKHS. First, these connections can inspire the design of new and more sophysticated regularizers: 
%examples have been provided in this paper
%and include the new kernels named iTC, iSS and iTS. 
%In addition, the Gaussian framework makes possible the use of effective statistical approaches for 
%estimating the hyperparameter vector. In fact, the marginal likelihood of $\eta$ is available in closed form  
%(differently from what happens e.g. when the Hankel nuclear norm is adopted), and this
%leads to a tuning approach whose favourable theoretical properties 
%have been also recently demonstrated in []ECC. 
 %in the intro cite Hankel by SIAM (Vanderberghe)

\section*{Appendix} 

\subsection{Hankel nuclear norm prior: approximation and MCMC reconstruction}
\label{App:Hankprior}
\label{PropAtom}
%\subsection{Prior induced by the Hankel nuclear norm: approximation and MCMC reconstruction}\label{App:Hankprior}

Our aim is to design an MCMC scheme able to reconstruct in sampled form
the prior 
$$
p_H(g)  \propto  \exp\left({-\frac{\sum_i \sigma_i(H)}{2\lambda}}\right)
%p_H(g) \propto \exp\left(-\frac{\|H\|_*}{\lambda}\right), \quad \|H\|_* = \sum_i \sigma_i(H) % = C e^{- \sum_i \sigma_i(H)} }
$$
associated to the Hankel nuclear norm regularizer. The key point is the definition of a 
proposal density leading to an efficient Metropolis-Hastings update, e.g. see \cite{Gilks}. 
For these purposes, it is useful to introduce the novel regularizer 
$$
J(g) = \sum_i \sigma^2_i(H).
$$
According to the Bayesian interpretation of
regularization, the associated prior is
\begin{align}\label{Htilde}
%\begin{center} p_H(g) \approx 
\tilde{p}_H(g) &\propto  \exp\left({-\frac{\sum_i \sigma_i^2(H)}{2\lambda}}\right) \\ \nonumber
&= \exp \left(-\frac{\Trace(HH^T)}{2\lambda}\right). 
%\end{center}
\end{align}
where $\Trace(HH^T)$ is the trace of $HH^T$.
The last equality, together with simple calculations, leads to the following result.

\begin{proposition}\label{tildep}
Let $g \in \R^m$ and $H \in \R^{p \times p}$, where $m=2p-1$. If $g$ is a random vector
with pdf  $\tilde{p}_H(g)$, then all the $g_k$ are independent and Gaussian. In particular, one has
\begin{eqnarray}\label{pHapprox}
g_k \sim \left\{ \begin{array}{cl}
    \mathcal{N}\left(0,\frac{\lambda}{k}\right)         & \   \mbox{if }  \          1 \leqslant  k  \leqslant  \frac{m+1}{2}  \\
    \mathcal{N}\left(0,\frac{\lambda}{m-k+1}\right) & \   \mbox{if }  \    \frac{m+1}{2} < k  \leqslant m  
\end{array} \right.
\end{eqnarray}
\begin{flushright}
$\blacksquare$
\end{flushright}
\end{proposition}

Thus, $\tilde{p}_H$ describes the impulse response coefficients as 
white noise whose variance first decreases until $k=\frac{m+1}{2}$, 
and then increases, a
stochastic process whose realizations are hardly similar to those of a stable system. 
Note also that the choice of the dimension $m$ of $g$ 
has an important influence on the prior shape as the minimum value is reached for $k=\frac{m+1}{2} $.\\ 
\color{black1} Coming back to the original Hankel prior,
we have exploited the prior $\tilde{p}_H$ to generate a Markov chain converging to $p_H(g)  \propto  \exp\left({-\frac{\sum_i \sigma_i(H)}{2\lambda}}\right)$ setting $2\lambda=1,g \in \R^{99},H \in \R^{50 \times 50}$. 
In particular, results displayed in Fig. \ref{Fig3} and discussed in subsection \ref{HankelPrior}, have been obtained
generating a chain of length 1e6 by a random walk Metropolis scheme. 
More specifically, when the state chain is $g^{k}$, the proposed sample is generated as 
$h^{k+1} = g^{k} + s^k$, 
where all the $s^k$ are i.i.d. random vectors drawn from $\tilde{p}_H$.
Then, with probability  $\min\left(1, p_H(h^{k+1})/p_H(g^{k})\right)$ %$\min\left(1, \frac{p_H(h^{k+1})}{p_H(g^{k})}\right)$
the Markov chain state $g^{k+1}$ is set to $h^{k+1}$, otherwise $g^{k+1}=g^k$.
\color{black} The left panel of
Fig. \ref{Fig3} also shows the standard deviations of the impulse response coefficients $g_k$ 
under the approximated prior $\tilde{p}_H$ characterized by
(\ref{Htilde}) (dashed line, scaled so that the variances of $g_1$ under $\tilde{p}_H$
and $p_H$ are equal). The
similarity between $\tilde{p}_H$ and $p_H$ confirms 
that the bad performance of the nuclear norm regularizer is due
to a prior which shares the same flaws pointed out by (\ref{pHapprox}).

\subsection{Proof of Proposition \ref{PropAtom}}\label{A2}

\color{black1}
We start  discussing the functional nature of the problem (\ref{MVatom2}) 
using RKHS theory, then obtaining its Bayesian interpretation.  
The main point of our proof 
is to exploit the RKHS representation in Theorem 4 on pag. 37 of \cite{Cucker01}
to connect the stable spline estimator with  
the atomic approaches.\\
Just for a while, it is useful to reason in continuous-time and 
introduce the following Sobolev space \cite{Adams} of functions $h:[0,1] \rightarrow \R$ 
\begin{eqnarray}\label{Sob}
\nonumber && \mathcal{S} = \left\{ h : h(0)=0, \ h \ \mbox{abs. cont.},  
\ \int_0^1 \dot{h}^2(t) dt < \infty \right\}
\end{eqnarray}
with (squared) norm $\| h \|_{\mathcal{S}}^2 = \int_0^1 \dot{h}(t)^2 dt$.  
It is well known that this is a RKHS with reproducing kernel 
which coincides with the covariance of the Brownian motion and is given, for $t,s \geq 0$ by 
\begin{equation}\label{Sexp}
S(s,t)=\min(s,t)=2\sum_{j=1}^\infty  \zeta_j \sin\left( \frac{t}{\sqrt{\zeta_j}}  \right) \sin\left( \frac{s}{\sqrt{\zeta_j}} \right)
\end{equation}
where $\zeta_j = \frac{1}{(j\pi - \pi/2)^2}. $
Combining (\ref{Sexp}) and RKHS theory \cite{Cucker01}, $\mathcal{S}$ can be also expressed as
\footnotesize 
\begin{equation}\label{Wrep} 
\mathcal{S}  =  \left\{
    h  \ | \   h(t) = \sum_{j=1}^\infty h_j \sqrt{2} \sin\left( \frac{t}{\sqrt{\zeta_j}}  \right) \ t \in [0,1], \
 \sum_{j=1}^\infty
    \frac{h_j^2}{\zeta_j } < \infty.
\right\}
\end{equation}
\normalsize
%with the following alternative formulation of the (squared) norm
%\begin{equation}\label{Wnorm}
%\| h \|_{\mathcal{S}}^2 = \sum_{j=1}^\infty \frac{h_j^2}{\zeta_j }  \left(= \int_0^1 \dot{h}^2(t) dt \right).
%\end{equation}

Now, note that $\min(\alpha^{t},\alpha^{s})=\alpha^{\max(t,s)}$. Then, still using
(\ref{Sexp}) we obtain% for the stable spline kernel (\ref{S1}):
\begin{equation}\label{Kexp}
\alpha^{\max(s,t)} = 2 \sum_{j=1}^\infty \zeta_j \sin\left( \frac{\alpha^{t}}{\sqrt{\zeta_j}}  \right) \sin\left( \frac{\alpha^{s}}{\sqrt{\zeta_j}} \right),
\end{equation}
which coincides with (\ref{ExpansionSS}) when $t$ and $s$ 
are restricted to the set of natural numbers $\mathbb{N}$. 
This expansion is key for our characterization. 
In fact,  according to (\ref{Kexp}) and Theorem 4 on pag. 37 of \cite{Cucker01}, the
RKHS induced by the stable spline kernel with domain on $\mathbb{N} \times \mathbb{N}$ contains the following functions $g: \mathbb{N} \rightarrow \R$:
\footnotesize 
\begin{equation}\label{Hrep}
\mathcal{H}  =  \left\{
    g   \ | \   g(t) = \sum_{j=1}^\infty g_j \sqrt{2} \sin\left( \frac{\alpha^{t}}{\sqrt{\zeta_j}}  \right) \ t \in \mathbb{N}, \
 \sum_{j=1}^\infty
    \frac{g_j^2}{\zeta_j } < \infty.
\right\}
\end{equation}
\normalsize 
But (\ref{Wrep}) and (\ref{Hrep}) also reveal that
$\mathcal{S}$ (which contains functions with domain $[0,1]$) and $\mathcal{H}$ 
(which contains functions of domain $\mathbb{N}$) share the same 
atomic expansion coefficients $g_j$ and $h_j$ and are 
isometrically isomorphic. 
%In particular, if $h(t) = g\left(-\frac{\log(t)}{\beta}\right)$, from (\ref{Wrep}-\ref{Hrep}) one has
%$$
%\| g \|_{\mathcal{H}}^2 = \int_0^{1} \dot{h}^2(t) dt = \int_0^{\infty} \dot{g}^2(t) \frac{e^{\beta t}}{\beta} dt.
%$$
%thus proving (\ref{JSS}).
In view of the RKHS connection, 
the solution of (\ref{MVatom2}) can be now obtained by 
the representer theorem for system identification.
More specifically, Theorem 3 on pag. 671 of \cite{SurveyKBsysid}
and the connection between 
Bayes estimation of Gaussian processes reported in Sections 1.4 and 1.5 of \cite{Wahba1990} 
lead to (\ref{MV}) and this completes the proof.
\color{black}

\bibliographystyle{plain}
\bibliography{biblioSurvey,biblio}

\end{document}